\pdfoutput=1 
\documentclass[12pt]{article}

\usepackage{inputenc}

\usepackage{amsmath,amssymb,bbm,color}
\usepackage{amsbsy}
\usepackage{euscript}
\usepackage{graphicx}
\usepackage{hyperref}
\usepackage{cite}
\usepackage{enumerate}
\numberwithin{equation}{section}

\newcommand{\afb}{{A_{FB}}}

\newcommand{\kksem}{{\tt  KKsem}}
\newcommand{\kkmc}{{\tt   KKMC}}
\newcommand{\kkfoam}{{\tt KKFoam}}

\newcommand{\bhlumi}{{\tt BHLUMI}}
\newcommand{\bhwide}{{\tt BHWIDE}}
\newcommand{\koralz}{{\tt KORALZ}}
\newcommand{\koralw}{{\tt KORALW}}
\newcommand{\korwan}{{\tt Korwan}}
\newcommand{\yfsww}{{\tt  YFSWW}}
\newcommand{\yfszz}{{\tt  YFSZZ}}
\newcommand{\kandy}{{\tt  KandY}}
\newcommand{\racoon}{{\tt RACOONWW}}
\newcommand{\wphact}{{\tt WPHACT}}
\newcommand{\winhac}{{\tt WINHAC}}
\newcommand{\photos}{{\tt PHOTOS}}
\newcommand{\tauola}{{\tt TAUOLA}}

\newcommand{\babayaga}{{\tt BabaYaga}}

\newcommand{\zfitter}{{\tt ZFITTER}}
\newcommand{\topazz}{{\tt  TOPAZ0}}

\newcommand{\order}[1]{${\cal O}(#1)$}

\begin{document}

\begin{titlepage}


\begin{flushright}
\bf IFJPAN-IV-2018-2
\end{flushright}

\vspace{5mm}
\begin{center}
    {\Large\bf QED challenges at FCC-ee precision measurements$^{\star}$ }
\end{center}

\vskip 5mm
\begin{center}
{\large S.\ Jadach and M.\ Skrzypek}

\vskip 2mm
{Institute of Nuclear Physics, Polish Academy of Sciences,\\
  ul.\ Radzikowskiego 152, 31-342 Krak\'ow, Poland}\\
\end{center}

\vspace{2mm}
\begin{abstract}
\noindent
The expected experimental precision of the rates and asymmetries
in the Future Circular Collider with electron positron beams (FCC-ee)
in the centre of the mass energy range 88-365GeV considered for construction in CERN,
will be better by a factor 5-200.
This will be thanks to very high luminosity,
factor up to $10^5$ higher than in the past LEP experiments.
Consequently, it poses the extraordinary challenge of improving the precision
of the Standard Model predictions by a comparable factor.
In particular the perturbative calculations of the trivial QED effects,
which have to be removed from the experimental data,
are considered to be a major challenge for almost all
quantities to be measured at FCC-ee.
The task of this paper is to summarize on the ``state of the art''
in this class of the calculations left over from the LEP era
and to examine what is to be done
to match the precision of the FCC-ee experiments --
what kind of technical advancements are necessary.
The above analysis will be done for most important observables
of the FCC-ee like the total cross sections near $Z$ and $WW$  threshold,
charge asymmetries, the invisible width of $Z$ boson, the spin asymmetry
from $\tau$ lepton decay and the luminosity measurement.
\end{abstract}

\vspace{50mm}
\footnoterule
\noindent
{\footnotesize
$^{\star}$This work is partly supported by
 the Polish National Science Center grant 2016/23/B/ST2/03927
 and the CERN FCC Design Study Programme.
}

\end{titlepage}

\newpage
\tableofcontents

\newpage
\section{Introduction}

The high-energy high-luminosity future circular electron-positron collider
FCC-ee~\cite{Gomez-Ceballos:2013zzn,Mangano:2018mur,Benedikt:2018qee},
considered for construction at CERN, 
would feature luminosity up to a factor $10^5$ higher than at LEP collider.
Together with the improvements of the detector techniques it would allow
to reduce an experimental error typically by a factor of 5-50,
and in some cases even by factor 200.
The uncertainty of the corrections due to trivial but large QED effects
to be removed from the data would then become for many observables
a dominant one and its reduction becomes a major challenge at the FCC-ee programme.
For instance, some QED effects of order 0.1\%, 
which at LEP could be neglected and accounted for in the error budget, 
will have to be calculated with two-digit precision, removed from data
and their uncertainty hopefully below 0.001\%
will enter into a combined experimental and theoretical error.
It is therefore important to review already 
now the present state of the art in the precise calculations
of the SM for $e^+e^-$ annihilation processes left over from the LEP era 
not only for multiloop pure electroweak 
corrections~\cite{Blondel:2018mad,Blondel:2019qlh,Blondel:2019vdq},
but also for QED effects
and to evaluate the prospects of the necessary future improvements 
in this area of the theoretical physics.

The modern techniques of calculating QED corrections in electron-positron
colliders for arbitrary experimental cut-offs using
Monte Carlo event generators were founded in 1980's
for in PETRA and PEP experiments.
Typically, these calculations were implementing complete \order{\alpha^1}.
Their precision tag (neglected higher orders) was of order 1-2\%,
see for instance ref.\cite{Berends:1980yz}.
Analytic phase space integration of QED distributions
was not playing at that time any major role,
apart from being used in designing Monte Carlo algorithms.
The LEP era has seen development of the entire new range of QED calculations for many
new processes, mostly in the form of the MC event generators.
However, for the analysing of data near the $Z$ resonance, numerical programs
based on analytic integration over the phase space were also playing an important role.
Also first QED calculations at the \order{\alpha^2} have appeared~\cite{Berends:1987ab}, 
but the biggest boost in the precision came from resummation to infinite order 
of soft photon corrections~\cite{Jadach:1988gb}
and from the calculations of the collinear logs due to small lepton masses
to higher orders, up to \order{\alpha^3}~\cite{Skrzypek:1992vk}.
Several of the MC programs were also armed with the 
complete \order{\alpha^1} electroweak (EW) corrections.
The example of the most sophisticated MC event generator from the LEP era
with soft photon resummation, complete \order{\alpha^2} QED and
\order{\alpha^1} EW corrections is \kkmc\ program~\cite{Jadach:1999vf}.
Generally, in LEP data analysis near the $Z$ peak, 
the use of MC generators was often limited
to removing detector inefficiencies and partly removing experimental cut-off effects.
Fitting EW parameters like masses of the top quark and the Higgs boson was done with
non-MC programs like \zfitter\cite{Bardin:1999yd} and 
\topazz\cite{Montagna:1998kp}.
However, at the LEP2 above the $WW$ threshold,
MC event generators~\cite{Jadach:2001mp,Denner:2002cg} were the only
tools capable to calculate QED+EW Standard Model predictions 
for the total cross section and distributions of the $e^+e^-\to W^+W^-$ process
and also were used to extract (fit) the mass of the $W$ boson from data.

As it will be argued in the following, due to significant
increase of the experimental and statistical precision at FCC-ee, 
the role of the MC event generators will increase.
Near $Z$ resonance a combination of MC programs and semianalytical non-MC calculations
will be used in the data analysis~\cite{Blondel:2018mad}%
\footnote{The most important role of semi-analytical calculations will be
   in testing/validating MC programs.}.
This will be due to many reasons, the rise of non-factorisable QED corrections
above the level of the experimental precision, 
the need of further development of the resummation techniques of soft and collinear
corrections which will be the only available within MC programs and the generally
bigger role of the multiparticle final states, for which the MC technique is
the only method of integration over the phase space.
The future tools and techniques 
for practical calculations of the QED effects will gradually emerge 
in more detail in the following discussion of many observables
to be measured at FCC-ee.

The content of the paper is the following:
we start with a brief overview of perturbative QED techniques.
In particular the soft and collinear factorisation 
is substantially  different from that of QCD. 
It will be stressed that optimal strategy
of truncation of the perturbative orders  from the point of view of precision
is not the simplistic order-by-order truncation but a more subtle approach
taking into account mass logarithms at higher orders.
Next we shall elaborate on the QED component of the systematic errors
in the measurements of the mass and partial width of $Z$, 
the total cross section and asymmetries near the $Z$ resonance 
and of the $W$ mass measurement,
according to the present state of art inherited from LEP era,
describing in detail the main source of 
QED uncertainties in these experimental observables.
Next, we shall compare present QED precision of the observables with
the planned experimental systematic and statistical precision at FCC-ee,
in order to estimate how much improvement is needed in the QED calculations.
For each experimental observable we shall examine how
difficult it will be to get sufficient improvement 
in terms of more perturbative orders, improvement in the resummation techniques
and development of new software tools i.e.
the entire new class of dedicated MC event generators.
In the above discussion we shall briefly elaborate upon the question
how to factorize in practice the resummed QED and the so called ``pure electroweak'' 
parts of the perturbative calculation such that they coexists in the MC
event generators forming a complete perturbative expansion without
any double counting nor under counting.
A short summary will complete the paper.

\section{General features of QED corrections}

\begin{figure}[t]
  \centering
  \includegraphics[height=65mm,width=165mm]{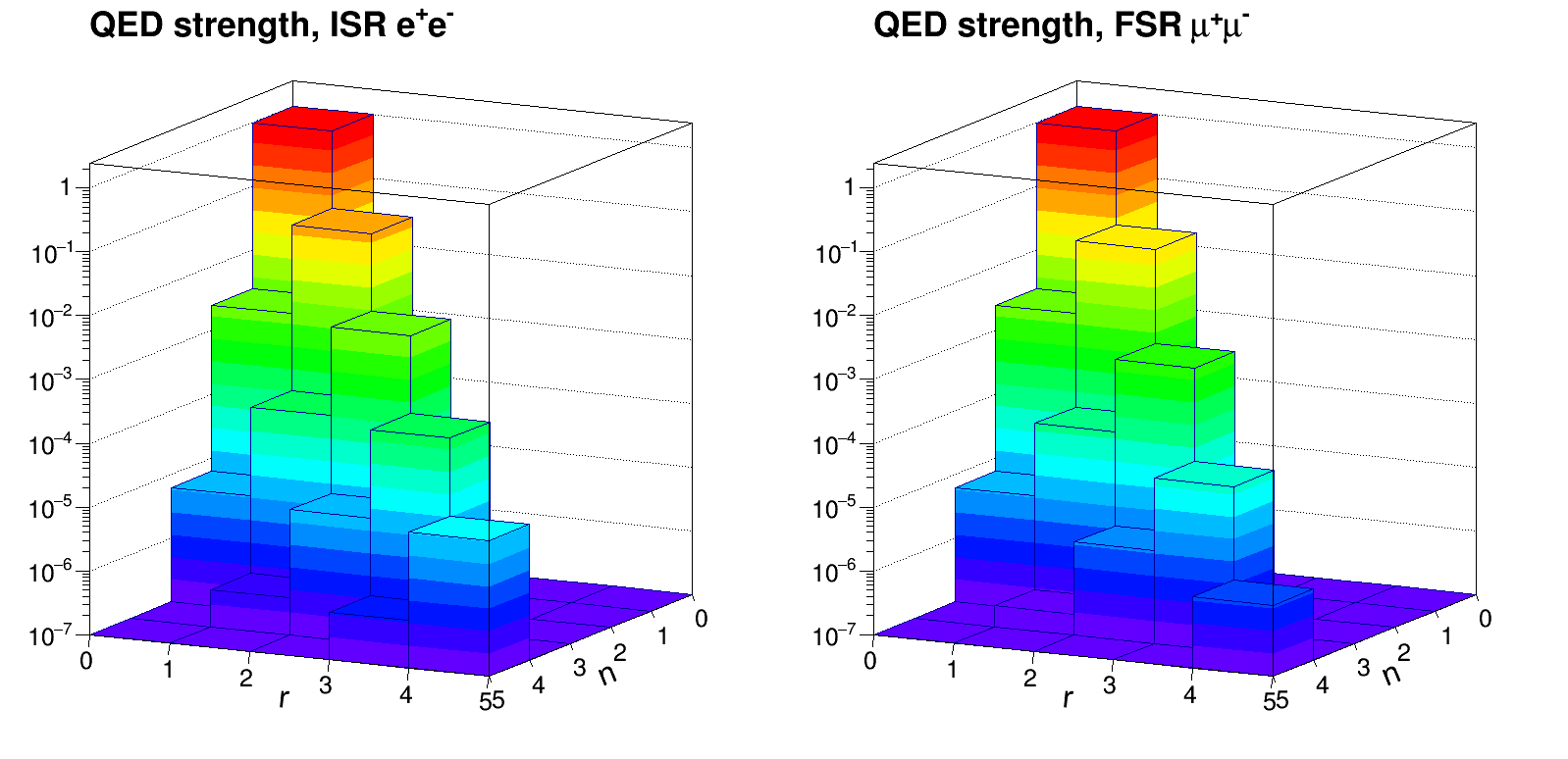} 
  \caption{\sf
  The parameter $\gamma_{nr}$ of eq.(\ref{eq:betanr})
  characterizing the size of the QED corrections.
  }
  \label{fig:cPrag2}
\end{figure}

\begin{figure}
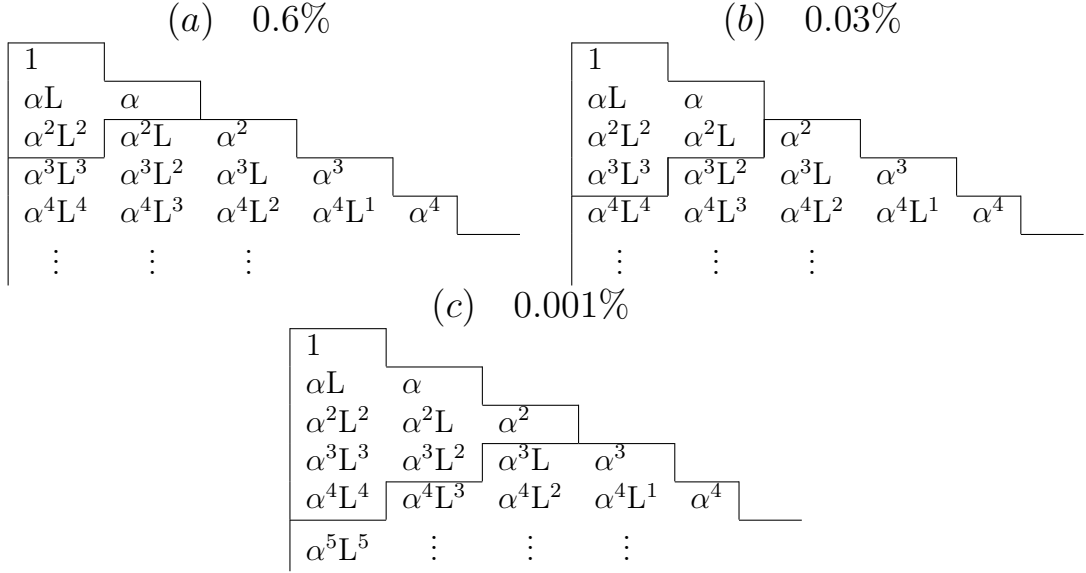

\centering
\begin{tabular}{cc}
\hbox{
\begin{tabular}{llllll}
& \multicolumn{3}{c}{\large $(a)\quad 0.6\%$}
\\
\cline{1-1}
\multicolumn{1}{|l|}{$1$}
\\ \cline{2-2}
\multicolumn{1}{|l}{$\alpha$L} &
\multicolumn{1}{l|}{$\alpha$}
\\ \cline{2-3}
\multicolumn{1}{|l|}{$\alpha^2$L$^2$} &
\multicolumn{1}{l}{$\alpha^2$L}       &
\multicolumn{1}{l|}{$\alpha^2$}
\\ \cline{1-1} \cline{4-4}
\multicolumn{1}{|l}{$\alpha^3$L$^3$}  &
\multicolumn{1}{l}{$\alpha^3$L$^2$}   &
\multicolumn{1}{l}{$\alpha^3$L}       &
\multicolumn{1}{l|}{$\alpha^3$}
\\ \cline{5-5}
\multicolumn{1}{|l}{$\alpha^4$L$^4$}  &
\multicolumn{1}{l}{$\alpha^4$L$^3$}   &
\multicolumn{1}{l}{$\alpha^4$L$^2$}   &
\multicolumn{1}{l}{$\alpha^4$L$^1$}   &
\multicolumn{1}{l|}{$\alpha^4$}       &
\hbox{\quad}
\\ \cline{6-6}
\multicolumn{1}{|c}{\vdots} &
\multicolumn{1}{c}{\vdots} &
\multicolumn{1}{c}{\vdots}
\end{tabular}}
\quad
\hbox{
\begin{tabular}{llllll}
& \multicolumn{3}{c}{\large $(b)\quad 0.03\% $}
\\
\cline{1-1}
\multicolumn{1}{|l|}{$1$}
\\ \cline{2-2}
\multicolumn{1}{|l}{$\alpha$L} &
\multicolumn{1}{l|}{$\alpha$}
\\ \cline{3-3}
\multicolumn{1}{|l}{$\alpha^2$L$^2$} &
\multicolumn{1}{l|}{$\alpha^2$L}     &
\multicolumn{1}{|l|}{$\alpha^2$}
\\ \cline{2-2} \cline{4-4}
\multicolumn{1}{|l|}{$\alpha^3$L$^3$} &
\multicolumn{1}{|l}{$\alpha^3$L$^2$}  &
\multicolumn{1}{l}{$\alpha^3$L}       &
\multicolumn{1}{l|}{$\alpha^3$}
\\ \cline{1-1} \cline{5-5}
\multicolumn{1}{|l}{$\alpha^4$L$^4$}  &
\multicolumn{1}{l}{$\alpha^4$L$^3$}   &
\multicolumn{1}{l}{$\alpha^4$L$^2$}   &
\multicolumn{1}{l}{$\alpha^4$L$^1$}   &
\multicolumn{1}{l|}{$\alpha^4$}       &
\hbox{\quad}
\\ \cline{6-6}
\multicolumn{1}{|c}{\vdots} &
\multicolumn{1}{c}{\vdots} &
\multicolumn{1}{c}{\vdots}
\end{tabular}}
\end{tabular}
%
\hbox{
\begin{tabular}{llllll}
& \multicolumn{3}{c}{\large $(c)\quad 0.001\% $}
\\
\cline{1-1}
\multicolumn{1}{|l|}{$1$}
\\ \cline{2-2}
\multicolumn{1}{|l}{$\alpha$L} &
\multicolumn{1}{l|}{$\alpha$}
\\ \cline{3-3}
\multicolumn{1}{|l}{$\alpha^2$L$^2$} &
\multicolumn{1}{l}{$\alpha^2$L}     &
\multicolumn{1}{l|}{$\alpha^2$}
\\ \cline{3-3} \cline{4-4}
\multicolumn{1}{|l}{$\alpha^3$L$^3$} &
\multicolumn{1}{l}{$\alpha^3$L$^2$}  &
\multicolumn{1}{|l}{$\alpha^3$L}       &
\multicolumn{1}{l|}{$\alpha^3$}
\\ \cline{2-2} \cline{5-5}
\multicolumn{1}{|l|}{$\alpha^4$L$^4$}  &
\multicolumn{1}{l}{$\alpha^4$L$^3$}   &
\multicolumn{1}{l}{$\alpha^4$L$^2$}   &
\multicolumn{1}{l}{$\alpha^4$L$^1$}   &
\multicolumn{1}{l|}{$\alpha^4$}       &
\hbox{\quad}
\\ \cline{1-1} \cline{6-6}
\multicolumn{1}{|c}{$\alpha^5$L$^5$} &
\multicolumn{1}{c}{\vdots} &
\multicolumn{1}{c}{\vdots} &
\multicolumn{1}{c}{\vdots}
\end{tabular}}
\caption{\sf QED perturbative leading and subleading corrections.
  Rows represent corrections in consecutive perturbative orders --
  the first row is the Born contribution.
  The first column represents the leading logarithmic (LO) approximation
  and the second column depicts the next-to-leading (NLO) approximation.
  In the figure, terms selected for the same precision level
  are limited with the help of an additional line.
}
\label{tab:pragma2}
\end{figure}

Fig.~\ref{fig:cPrag2} shows parameters which control
effectively the strength of the QED ${\cal O}(\alpha^n)$
corrections for $e^\pm$ beams (ISR)
and for final state muon pair $\mu^\pm$ (FSR) at the $Z$ peak:
\begin{equation}
\label{eq:betanr}
 \gamma_{nr}=
 \bigg( \frac{\alpha}{\pi} \bigg)^n  
 \bigg( 2 \ln\frac{M_Z^2}{m_f^2} \bigg)^r,\; 0\leq r \leq n,
\end{equation}
for $f=e,\mu$ correspondingly.
The relative precision $6\cdot 10^{-3}$ requires the inclusion of corrections of
the QED correction up to
\[
 {\cal O}( \alpha^1 L_f^1,  \alpha^1 L_f^0,\alpha^2 L_f^2 ),\quad
 L_f= \ln(s/m_f^2),
\]
while the remaining second order correction 
$ {\cal O}(\alpha^2 L_f^1,\alpha^2 L_f^0)$ 
can be safely neglected.
This is also visualized in Fig.~\ref{tab:pragma2}(a).

The next precision level  of $3\cdot 10^{-4}$
requires adding 
$ {\cal O}(\alpha^3 L_f^3,\alpha^2 L_f^1 )$,
see also Fig.~\ref{tab:pragma2}(b) for the illustration.
It is only at the  precision level  $1\cdot 10^{-5}$
(relevant for the FCC-ee)
that including $ {\cal O}(\alpha^2L_f^0)$ looks necessary,
but it should to be accompanied in addition with
$ {\cal O}(\alpha^4 L_f^4,\alpha^3 L_f^2 )$
corrections in order to be useful in practice.

The case of muon pair differs only slightly,
as can be read from the RHS plot in  Fig.~\ref{fig:cPrag2}.

\subsection{Optimal strategy for advancing QED precision}

In view of the above estimates of the numerical importance 
of the generic perturbative QED contributions
it is easy to answer the question about
the optimal strategy for advancing the precision of QED calculations.
For example, performing the complete $ {\cal O}(\alpha^3)$
calculation may look justified because it will be automatically
gauge invariant, all soft/collinear limits, mass effects will be correct.
However, for precision at the $\sim 0.5\%$ level,
a much simpler (almost trivial) calculation
taking only up to \order{\alpha^2 L^2_f} into account is enough, 
see Fig.~\ref{fig:cPrag2}.
Similarly, the inclusion of $ {\cal O}(\alpha^2 L^1_f)$
in order to gain precision $\sim 0.05\%$ level is in
practice useless, unless $ {\cal O}(\alpha^3 L^3_f)$ contribution
enters into the game, see Fig.~\ref{fig:cPrag2}.

It is also a common misconception that including $ {\cal O}(\alpha^2 L^2_f)$
forces us into the use of the strictly collinear approximation%
\footnote{%
   By ``strictly collinear'' we mean that in the structure function 
   photon angle is integrated over, i.e. photon is treated partly inclusively.}.
It was demonstrated that in the EEX matrix element\cite{Jadach:2000ir} 
with \order{\alpha^2 L^2_f,\alpha^3 L^3_f} corrections
can be implemented at the exclusive (unintegrated) level,
i.e. keeping angle distributions for two and more photons
accessible for experimental cut-offs.

\subsection{More on soft and collinear resummation in QED calculations}

The other important question is that 
of the importance of the soft and collinear photon resummation.
For totally inclusive observables in the final state 
(like FSR in the $Z$ boson decay)
mass logarithms do cancel completely among real and virtual ones
according to the Kinoshita-Lee-Nauenberg theorem.
For the cut-off on the total photon energy $E_{\max} \ll E_{beam}$,
the $ {\cal O}(\alpha^n)$ YFS/Sudakov double logarithmic contributions
\[
   S^{(n)} \sim \Big( \frac{\alpha}{\pi}\; 2L_f\;
               \ln\frac{E_{\max} }{E_{beam}} \Big)^n
\]
for a strong cutoff may easily be huge, $ S^{(n)} \gg 1$,
and definitely require resummation.
Only for a loose cut-off like $ E_{\max}/E_{beam} \sim 0.2$
and very low precision requirements 
one could avoid soft photon resummation.

There are several techniques
of soft photon resummation of the different levels of sophistication.
In the most primitive version all photon energies are restricted to small
values, the integration over photon angles and
energies is done analytically, keeping only the sum of photon energies fixed.
This we call {\em inclusive soft photon exponentiation}, IEX in short.
In the Yennie-Frautschi-Suura (YFS) work~\cite{Yennie:1961ad} it was outlined 
how to match smoothly soft and hard photon distributions,
covering the entire phase space, without any need of the cut-off
on total photon energy.
It was only in late 1980's that it was gradually worked out
how to implement YFS soft photon factorisation and resummation
within the Monte Carlo event 
generators~\cite{Jadach-yfs-mpi:1987,Jadach:1988rr}.
This technique we call {\em exclusive photon exponentiation}, EEX in short.
In the EEX methodology soft photon factorisation and resummation
is done at the level of the multiphoton fully differential distributions,
that is for spin amplitudes squared and spin summed/averaged.

Near the narrow resonances $R$ another similar class of soft logarithms
\[
   S^{(n)} \sim \Big( \frac{\alpha}{\pi}\;
               \ln\frac{\Gamma_R }{M_R} \Big)^n
\]
present in real and virtual corrections also requires resummation.
In addition, due to the complicated pattern of the QED interferences
between the initial and final state photons
(cancellations due to short lifetime of the resonance)
which operate at the amplitude level,
one is forced to perform soft photon factorisation at the amplitude level,
before squaring and spin summing.
This technique was developed by the Frascati group~\cite{Greco:1975rm}.
Moreover, the need of proper implementation of spin effects
in the $\tau$ pair production and decays at LEP also was enforcing
the use of spin amplitudes.
In order to meet the above requirements a new variant of YFS-inspired
soft photon resummation was developed~\cite{Jadach:1998jb}, 
in which soft photon {\em factorisation} was reformulated at the amplitude level
and soft photon {\em resummation} is implemented numerically within the MC program.
Narrow resonance effects were accommodated as in 
ref.~\cite{Greco:1975rm,Greco:1980mh}.
This technique we call {\em coherent exclusive exponentiation}, CEEX in short.
So far, the only implementation of CEEX technique is 
in the \kkmc\ event generator~\cite{Jadach:1999vf}.

The important message to theorists specialising in QED+EW multiloop 
calculations is the following:
{\em do not add soft real emissions to multiloop results in order to
eliminate infrared singularities \'a la Bloch-Nordsieck}, 
if you want these results to be used
in the MC generators with IR resummation.
Instead, you should subtract IR parts (YFS virtual formfactor) from the amplitudes,
before squaring and spin summing%
\footnote{One can do it for gauge invariant groups of diagrams
   but not for individual diagrams. 
   Undoing Bloch-Nordsieck at \order{\alpha^2} is usually unfeasible,
   rendering two-loop calculations useless for the MC.}.
Why? Because combining IR soft and real contributios and the differential
cross section level is already done in the Monte Carlo.

The related important practical question is whether the use of collinear resummation
of the mass logarithms is mandatory in QED to infinite order, like in QCD?
Obviously in QED it is not mandatory and
in practice it is usually enough to stop at some finite order, 
typically truncating infinite LO series at $ {\cal O}(\alpha^3 L^3_f)$.
In the FCC-ee environment it may be sometimes
necessary to include $ {\cal O}(\alpha^4 L^4_f)$.

In case of the photon emission from leptons,
employing the entire machinery of the collinear resummation 
technique up to LO+NLO or LO+NLO+NNLO level 
is not trivial due to finite lepton masses, handling properly
the factorisation scale parameter, sectorisation of the phase space etc.
In practice, it is more convenient and/or economical to
perform soft photon resummation in first place and only then
to include collinear resummation truncated to a convenient order%
\footnote{%
 Due to the more complicated infrared limit and almost zero quark masses, 
 resummation in QCD is done in different order --
 collinear resummation and soft resummation are in principle done simultaneously,
 with soft limit approximated quite often to some convenient order within the 
 leading-logarithmic expansion.
}.
This technique was used successfully in the many MC projects like 
\koralz\cite{Jadach:1993yv},
\kkmc\cite{Jadach:1999vf}, 
\bhlumi\cite{Jadach:1996is}, 
\yfsww\cite{Jadach:2001uu}, 
\koralw\cite{Jadach:1998gi}, 
\bhwide\cite{Jadach:1995nk}, 
\yfszz\cite{Jadach:1996hh},
also in the semianalytic approaches like 
\zfitter~\cite{Bardin:1999yd} and \topazz~\cite{Montagna:1998kp}.
Moreover, soft photon resummation is of the highest possible
priority for the resonant processes, obviously near the $Z$ resonance, 
but also it will be mandatory in the $WW$ production process,
especially near the $WW$ threshold.

Summarizing the above discussion:
\begin{itemize}
\item
The (complete) order-by-order perturbative calculation in QED is definitely not
the economic way to to obtain predictions for cross sections or
asymmetries with the precision below 0.1\%.
\item
Soft photon resummation is an absolute necessity, especially for resonant processes.
It exists in at least three different variants (IEX, EEX and CEEX).
\item
Do not follow Bloch-Nordsieck to eliminate IR singularities!
\item
Resummation of collinear mass logarithms $\ln(s/m_f^2)$ is very useful, 
but in QED it is usually convenient to truncate it at some finite order.
\item
Approximation of small lepton mass $m_f^2/s \ll 1$ should be exploited
for electron and muons as much as possible, 
but for $\tau$ lepton
$\propto m_\tau^2/s, m_\tau^4/s^2$ terms
may not be negligible at tree-level, while
higher powers of this type in higher orders of $\alpha$ are probably irrelevant.
\end{itemize}

\subsection{QED ``deconvolution''}

The so-called ``deconvolution'' of QED effects is the procedure of
removing universal (process independent) QED effects from experimental
data for the total cross section and angular differential cross sections
in the process $e^+e^-\to f\bar{f}(+n\gamma)$.
It was a cornerstone of the final analysis of LEP data
near the $Z$ peak in ref.~\cite {ALEPH:2005ab}.
Cross sections, asymmetries, branching ratios, $Z$ mass and partial widths
derived from  the ``deconvoluted'' integrated/differential experimental
distributions
were called EW pseudo-observables (EWPOs) 
and were (approximately) independent of the QED effects and experimental cutoffs,
such that they could be combined among four LEP collaboration and SLD.
The technique of EWPOs was
defined and thoroughly elaborated in ref.~\cite{Bardin:1999gt}.

N.B. The term ``deconvolution'' is a little bit misleading, because in practice
it means fitting a certain theoretical formula to experimental data
at one or several energies%
\footnote{The true deconvolution 
would require taking data for $d\sigma(s)/d\cos\theta$
on a dense grid of $s$ and $\theta$ and calculating the deconvoluted
differential distribution at $s$ by means of combining all data $s'<s$,
using some weight provided by QED, without any fitting.}.

It is tempting to assume
that at FCC-ee one can use the same QED deconvolution as at LEP.
However, this assumption may be wrong
due to the much higher experimental precision at FCC-ee.

Let us comment briefly on some aspects of 
{\em the factorisation of the soft and collinear QED corrections}, 
which is the basis of QED deconvolution
and an essential element in the construction of EWPOs.
This is because one could worry that
the entanglement of electromagnetic and weak interactions at multiloop corrections,
the rise of non-factorisable interferences above the level of FCC-ee precision,
may give rise to practical or principle problems with the clean separation of 
the QED universal corrections from the complete EW perturbative calculations, 
especially beyond the 1-st order.

In refs.~\cite{ALEPH:2005ab,Bardin:1999gt} 
the simple version of QED ISR deconvolution relies on a well
known simple convolution formula with the integration over a {\em single}
variable (total ISR photon energy) over the product of two objects:
the ISR radiator function%
\footnote{The ISR radiator function was that of ref.~\cite{Jadach:1999pp}
 with \order{\alpha^1L_e,\alpha^1,\alpha^2L_e^2,\alpha^2L_e,\alpha^3L_e^3} 
 photonic corrections, contributions from soft fermion pairs and soft photon resummation.
}
and Born-like differential or integrated cross section.
In the case of a cut-off on the final fermion pair effective mass, 
the integration over the radiator function of FSR had to be included.
The resulting double convolution formula for ISR$\times$FSR was used
in ref.~\cite{Bardin:1999gt} and also in ref.~\cite{Jadach:2000ir} for testing 
\kkmc.
In addition, in the \zfitter\ and \topazz\ programs missing \order{\alpha^1} 
contributions were added into the game.
In particular, missing initial-final state interference (IFI) 
and effects due to experimental cut-offs were also added in this way.

As it was discussed in ref.~\cite{Bardin:1999gt}, the above simple treatment
of IFI at LEP analysis was possible because near the $Z$ peak, 
in the absence of strong cutoffs IFI is suppressed 
by an additional $\Gamma_Z/M_Z$ factor,
while away from the $Z$ peak, where LEP data were of limited precision 
and IFI also could be either neglected or eliminated using 
an additive \order{\alpha^1} correction.
The above treatment of IFI would be highly unsatisfactory 
at the FCC-ee precision.

The first mandatory thing on the way to improved treatment of IFI is soft
photon resummation.
The multiphoton convolution formula at the level of the matrix element 
in the soft photon approximation was constructed
a long time before the LEP era, by the Frascati group
\cite{Greco:1975rm,Greco:1975ke}, and it is a natural extension
of the the exponentiation formula of 
Yennie-Frautschi-Suura~\cite{Yennie:1961ad} to a resonant process.

In the Frascati-type formula with soft photon resummation
{\em at the amplitude level} the effective Born matrix element is clearly
factorized out and could be exploited for constructing a better variant
of the EWPO definition.
However, as it was shown in ref.~\cite{Jadach:2018lwm},
after squaring, spin-summing,
and integrating over photon angles the resulting convolution formula,  
has four convolution variables, 
two for ISR and FSR and two additional variables for IFI.
It is not so handy as the traditional one,
because the 5-dimensional (including $\cos\theta$)
integration has to be done numerically using the MC method.
In ref.~\cite{Jadach:2018lwm}
new MC code \kkfoam\ implementing the above calculation was used to crosscheck
the calculation of the \kkmc\ program in the soft limit, where it should be
by construction fully compatible with the Frascati approach.

In the final LEP data analysis~\cite {ALEPH:2005ab}
the coupling constants of $Z$, its mass and width
inside the effective Born (differential) cross section
were obtained from the fit to data taken typically only at 3-5 energies.
The deconvoluted Born integrated cross section and asymmetries were
not really coming directly from data, but were calculated
from the (theoretical) fitted effective Born at $s=M_Z^2$.
In other words, the resulting EW pseudo-observables (EPWOs) were encoded in the 
parameters inside the effective Born, EW pseudo-parameters (EWPPs).
The definition of the effective Born in ref.~\cite {ALEPH:2005ab}
was done at the spin amplitude level, see eq. (1.34) therein.
Could the above scenario be repeated at FCC-ee using an effective Born
defined at the amplitude level and factorizing QED at the amplitude level?
The detailed numerical studies, as in ref.~\cite{Bardin:1999gt},
at the FCC-ee precision level have to be done from the scratch
in order to answer this question.
In particular an additional uncertainty introduced 
by partial/incomplete inclusion of the 
SM effects in EWPOs extracted from data should be re-examined.
In the recent ref.~\cite{Blondel:2018mad} it was argued that the combined use
of  more advanced versions of the \zfitter/\topazz\ programs
and of the MC programs of the \kkmc\ class may provide solution.

Summarizing: At the FCC-ee precision IFI requires resummation.
Clean factorisation into a 
and a model-independent (Born-like) part
still works at the spin amplitude level, 
even if it fails (due to non-factorisable contributions)
at the amplitude squared level.
Such a factorisation opens ways to a new more flexible
definition of EW pseudo-observables, 
which would possibly cope with the FCC-ee precision,
see subsection~\ref{ssect:ewpos} for more details.

\section{The most important QED-sensitive experimental observables at LEP and FCC-ee}

\begin{table}
\centering
\begin{tabular}{|c|c|c|c|c|c|}
\hline
 Observable & Where from & Present (LEP) & FCC stat. & FCC syst& 
$\frac{\rm Now}{\rm FCC}$\\
\hline
$M_Z$ [MeV]    & $Z$ linesh.\cite{ALEPH:2005ab}
               & $91187.5\pm 2.1\{0.3\}$ & $0.005$ & $0.1$ 
               & 3 \\
$\Gamma_Z$ [MeV] & $Z$ linesh.\cite{ALEPH:2005ab}
               &  $2495.2\pm 2.1\{0.2\}$ & $0.008$ & $0.1$ 
               & 2\\
$R^Z_l=\Gamma_h/\Gamma_l$ & $\sigma(M_Z)$\cite{Abbaneo:2001ix}
               & $20.767\pm 0.025\{0.012\}$& $ 6\cdot 10^{-5}$&$ 1\cdot 10^{-3}$
               & 12\\
$\sigma^0_{\rm had}$[nb] & $\sigma^0_{\rm had}$ \cite{ALEPH:2005ab}  
               & $41.541 \pm 0.037\{0.025 \}$
               & $ 0.1\cdot 10^{-3}$ & $ 4\cdot 10^{-3}$ 
               & 6\\
$N_\nu$        & $\sigma(M_Z)$\cite{ALEPH:2005ab}
               & $2.984\pm 0.008\{0.006\} $  &$ 5\cdot 10^{-6}$ &$1 \cdot 
10^{-3}$
               & 6\\
$N_\nu$        & $Z\gamma$ \cite{Abbiendi:2000hh}
               & $2.69\pm0.15\{0.06\} $ & $ 0.8 \cdot 10^{-3}$& $<10^{-3}$ 
               & 60\\
$\sin^2\theta_W^{eff}\!\times 10^{5}$ 
               & $A_{FB}^{lept.}$\cite{Abbaneo:2001ix}
               & $23099\pm 53\{28\}$ & $ 0.3$ &  $ 0.5$ 
               & 55\\
$\sin^2\theta_W^{eff}\!\times 10^{5}$ 
               &$\langle{\cal P}_\tau\rangle$,$A_{\rm 
FB}^{pol,\tau}$\!\!\cite{ALEPH:2005ab} 
               & $23159\pm 41\{12\}$ & $0.6$ & $<0.6$ 
               & 20 \\
$M_W$ [MeV]    & ADLO\cite{Schael:2013ita}
               & $80376\pm 33\{6\}$  &   0.5    &  0.3   
               & 12 \\
 {\small $A_{FB,\mu}^{M_Z\pm 3.5 {\rm GeV}}$}
 & $\frac{d\sigma}{d\cos\theta}$\cite{ALEPH:2005ab}
               & $\pm 0.020\{0.001\}$
               & $1.0\cdot 10^{-5}$  & $0.3\cdot 10^{-5}$  
               & 100 \\
\hline
\end{tabular}
\caption{\sf
 Listed are electroweak observables, 
 which are most sensitive to QED effects.
 Experimental (LEP) errors in the 3-rd column are accompanied with error component
 in the braces $\{...\}$ {\em induced} by QED calculation uncertainties.
 FCC-ee experimental systematic errors in 4-th column are
 from FCC-ee CDR~\cite{Mangano:2018mur}
 except $\tau$ polarisation~\cite{2019:cern-talk-tenchini}.
 They are all without theoretical uncertainty component.
 Last column shows improvement factor in QED theoretical calculations
 needed in order to be equal to experimental precision of FCC-ee measurements.
}
\label{tab:table1}
\end{table}

The minimum {\em improvement factor}
to be achieved in the precision of QED perturbative calculations for FCC-ee experiments,
in order that these effects are controlled at the level of the pure experimental errors,
are shown in the last column of Table~\ref{tab:table1}.
In the third column there we collect 
the values of the total error of the observables
most sensitive to QED effects, as measured at LEP,
including also explicitly QED theoretical uncertainty induced 
in the overall experimental systematic errors.
These components are accompanied with the citations of the source papers
from which they are taken.
The projected much better 
experimental statistical and systematic errors at FCC-ee experiments
are shown in 4th ad 5th column.
They are taken from Table~S.3 in ref.~\cite{Mangano:2018mur}.
As a matter of fact, the actual {\em improvement factor} for FCC-ee
should be even 2-3 times bigger
than the one shown in Table~\ref{tab:table1},
in order to be sure that QED effects are clearly a subdominant component
in the corresponding overall systematic error
for most of the observables measured at LEP experiments.

The information in the Table~\ref{tab:table1} is the starting point
for more detailed discussion in the following sections.
The most sensitive to QED effects observables of the FCC-ee experiments
listed in the table are total cross section,
including low angle Bhabha for luminosity measurement,
especially near the $Z$ resonance (for $Z$ mass and width, $Z$ invisible width),
cross section of $Z\gamma$ final state ($Z$ radiative return) above the $Z$ peak,
charge asymmetry for leptonic pair final states
and spin asymmetry in the $\tau$ pair production.
We are analysing a subset of experimental observables near the $Z$ resonance
and $WW$ threshold which are most ''vulnerable'' to the QED effects --
omitting many others.

Let us comment briefly on Table~\ref{tab:table1},
before more detailed discussion in the following sections:
QED uncertainty of $M_Z$ and $\Gamma_Z$ (derived from the lineshape)
are taken from refs.\cite{ALEPH:2005ab,Jadach:1999gz}.
Huge photonic corrections to the $Z$ lineshape (30\%) in all LEP experimental
and theoretical studies are mastered using a formula
derived in ref.~\cite{Jadach:1992aa}.
However, the precision of QED corrections in ref.\cite{ALEPH:2005ab}
is dominated by the uncertainty of the fermion pair correction
of ref.~\cite{Arbuzov:2002rp},
in spite of the fact that this kind of correction is small by itself.
For more discussion on that see sect.~\ref{sec:lineshape}.
The QED uncertainty of $R^l_Z$  is taken from \cite{Abbaneo:2001ix} 
and is mainly due to $t$-channel%
  \footnote{%
   The $t$-channel subtraction \cite{ALEPH:2005ab}
   done using ALIBABA program \cite{Beenakker:1990mb} 
   instead of more sophisticated
   BHWIDE Monte Carlo \cite{Jadach:1995nk} enhanced
   unnecessarily this problem. }
obscuring $\Gamma_{ee}$.
QED uncertainty of $N_\nu$ from the $Z$ radiative return is from
the OPAL paper \cite{Abbiendi:2000hh} and its error is enhanced
due to the use of the outdated MC programs%
\footnote{\koralz\ and {\tt NUNUGPV98} were used instead of \kkmc.}
with inferior precision $\delta \sigma /\sigma =2\%$.
The factor 60 in last column is therefore a gross overestimate.
On the other hand the precision of
$N_\nu$ from the peak cross section is limited by the luminosity error
$\delta {\cal L}/{\cal L} \simeq 0.6 \cdot 10^{-3}$,
mainly due to QED corrections for the low angle Bhabha luminometer%
\footnote{ The LEP era approximate rule for the luminosity error contribution
  to the error of the number of neutrino species  
  was $\delta N_\nu \simeq 7.5~\delta {\cal L}/{\cal L}$.
}.
Strictly speaking this luminosity error at LEP was mainly of the QCD origin
because it was dominated by the vacuum polarisation contribution uncertainty
due to the experimental error of low energy hadronic data.

The leptonic charge asymmetry of LEP experiments $A_{FB}^{lept.} = 0.0171(10)$ 
from Table 2.13 in ref.~\cite{ALEPH:2005ab}
translates into EW mixing angle $\sin^2\theta^{eff}_W = 0.23099(53)$,
see Table 12.4 in ref.~\cite{Abbaneo:2001ix}).
It is almost the same as for the muon pair final state alone.
The theoretical error component was estimated 
in ref.~\cite{ALEPH:2005ab} (Table 2.8) as 
$\delta A_{FB}^{0,l} = 0.01\%$ from the difference between \zfitter\ and
\topazz\,
see also ref.~\cite{Bardin:1999gt}.
This difference provides for the technical precision of both programs
and not what we need, that is uncertainty due to missing higher orders.
Such an estimation of the QED perturbative uncertainty of $\afb$ 
due to missing higher orders was provided in ref.~\cite{Jadach:2000ir}.
From the comparison of the results of $\afb(M_Z)$ 
from \kkmc, \order{\alpha^1} \koralz\ and \zfitter\ (IFI included)
in Fig.~28(b) for the loose cut-off on total photon energy
the uncertainty of order $\delta\afb \simeq 0.05\%$ seems reasonable.
It translates into $\delta \sin^2\theta^{eff}_W = 0.57\cdot \delta A_{FB}^{0,l} 
\simeq 0.00027$.

The effective $\sin^2\theta^{eff}_W = 0.23159(41)$ in Table~\ref{tab:table1}
from two spin asymmetries,
$\langle {\cal P}_\tau \rangle$ and $A_{\rm FB}^{pol,\tau}$,
comes from the LEP summary of ref.~\cite{ALEPH:2005ab} (Section 4.4).
Both asymmetries have comparable experimental errors.
QED effects in the MC simulation of the $\tau$ decays 
quoted there following ref.~\cite{Barberio:1990ms} are estimated to contribute
$\delta {\cal A}_\tau=0.0010$ for $\langle {\cal P}_\tau \rangle$,
which yields $\delta \sin^2\theta^{eff}_W = (1/8)\delta {\cal A}_\tau= 0.00012$.
For $A_{\rm FB}^{pol,\tau}$ the QED uncertainty according to 
ref.~\cite{ALEPH:2005ab} is even smaller.
The corresponding anticipated FCC-ee experimental statistical and systematic
errors are taken from ref.~\cite{Mangano:2018mur}, 
where it is shown for $A_{\rm FB}^{pol,\tau}$.
We transformed it into the error of the EW mixing angle using relation
$\delta A_{\rm FB}^{pol,\tau} \simeq 6\cdot \delta \sin^2\theta^{eff}_W $.

The combined LEP measurement of $M_W$ in ref.~\cite{Schael:2013ita}
comes mainly from the $W$ mass reconstruction in the $q\bar{q}q\bar{q}$
and $l\nu_l q\bar{q}$ channels.
The uncertainty of 6MeV due to higher order radiative corrections is estimated
from comparison of the \kandy\cite{Jadach:2001mp,Jadach:2000kw}
and \racoon\cite{Denner:2002cg} MC programs.

The LEP experimental error of $A_{FB,\mu}^{M_Z\pm 3.5 {\rm GeV}}$ 
being $\delta A_{FB}\sim 2\%$ is mainly statistical,
and the QED error estimate $\delta A_{FB}\sim 0.1\%$
is taken from refs.\cite{Jadach:2000ir,Kobel:2000aw}.

Summarizing on the Table~\ref{tab:table1} one may say that the very minimum
of the improvement on the QED calculations 
needed for FCC-ee near the $Z$ resonance
and near the $WW$ threshold is typically a factor $\sim$ 3--60, 
with the exception of charge asymmetry, 
where a bigger improvement $\sim 100$ will be necessary.
It is, however, important to stress that in order to get to the same
comfortable situation as at LEP, where QED uncertainties 
(except for the luminosity cross section) were typically at least factor 2-3
smaller than the statistical and systematic experimental error,
one should really aim at the improvement factors being $\sim 3$ times bigger
than these of the last column in Table~\ref{tab:table1},
that is a factor 10-180 and in the special case of $\afb$ a factor 300.
With values shown in the last column in Table~\ref{tab:table1}
one will merely get into a rather uncomfortable situation like in the measurement
of the luminosity and the total cross section at LEP1, for which theory uncertainty
was comparable to the experimental error or even slightly bigger.

\section{Present state of the art in calculating QED effects}
In the following we are going to elaborate in a more detail
on the present state of art in calculating of the QED effects,
as in most cases inherited from the LEP era.
It will be done for each observable listed in the tab.~\ref{tab:table1}.

\subsection{Cross section near the $Z$ resonance -- mass and width of $Z$}
\label{sec:lineshape}
The most sizable QED corrections to the total cross section near the $Z$ resonance
i.e. the so called {\em lineshape} ($\sim$ 30\%), 
are due to multiple photon emissions from the initial $e^\pm$ beams (ISR).
The basic ``effective radiator'' formula for the ISR {\em photonic} QED corrections
was defined in ref.\cite{Jadach:1992aa},
combining/summarizing results of several other works.

The master ISR formula of ref.\cite{Jadach:1992aa} includes,
in addition to the classic ${\cal O}(\alpha^1)$ result\cite{Bonneau:1971mk},
the relatively simple ${\cal O}(\alpha^2 L_e^2)$ contribution \cite{Altarelli:1986kq}
and the more elaborate ${\cal O}(\alpha^3 L_e^3)$ corrections 
obtained in the analytical form%
\footnote{%
  This result was later on correctly reproduced in ref.~\cite{Montagna:1996jv}. 
}
in ref.\cite{Jadach:1990vz} and
crosschecked using dedicated MC program\cite{Skrzypek:1992vk}.
The next correction ${\cal O}(\alpha^4 L_e^4)$ is also available,
see ref.\cite{Przybycien:1992qe}.
According  to analysis of ref.\cite{Jadach:1999pp},
at the LEP precision it is negligible.
ISR formula of ref.\cite{Jadach:1992aa}
also includes  ${\cal O}(\alpha^2 L_e^1)$ photonic contribution 
from ref.\cite{Berends:1987ab},
see alse refs.~\cite{Blumlein:2019srk,Blumlein:2011mi} 
for corrected  ${\cal O}(\alpha^2 L_e^0)$ results.
In addition to photonic corrections, it includes also the small contribution 
from fermion pair production of ref.\cite{Skrzypek:1992vk}.
It should be remembered that a similar but less precise ISR formula
was presented earlier in the pioneering work of ref.~\cite{Kuraev:1985hb}.
Both ISR formulas of refs.\cite{Jadach:1992aa,Kuraev:1985hb} are
employing soft photon resummation%
\footnote{%
  YFS resummation of ref.\cite{Jadach:1992aa} has turned out to be more 
effective in resumming higher orders.}.

The master analytical formula for ISR of ref.\cite{Jadach:1992aa}
was used in the numerical studies of the $Z$ lineshape ($|\sqrt{s}-M_Z|<3$GeV)
in refs.~\cite{Jadach:1999pp,Jadach:1999gz} and 
in the analysis of all LEP experimental data
and all related theoretical studies,
as a basic tool for evaluating QED effects in the $Z$ lineshape.
In ref.~\cite{Jadach:1999pp} the total QED uncertainty due
to photonic corrections was estimated to be
$\delta M_Z, \delta \Gamma_Z\leq 0.1$MeV.

The master convolution formula of ref.\cite{Jadach:1992aa}
with the ISR radiator function
was neglecting contribution from the initial-final state interference (IFI).
The uncertainty due to IFI to the lineshape observables was very carefully
analyzed in ref.\cite{Jadach:1999gz},
exploiting older ${\cal O}(\alpha^1)$ analytical work 
(without exponentiation) \cite{Jadach:1987ws},
and results from MC programs with~\cite{Jadach:1999vf} 
and without~\cite{Jadach:1993yv} soft photon resummation.
The IFI correction was found to be of the order
$\delta\sigma/\sigma \sim 0.2\%$ for $|\sqrt{s}-M_Z|\leq 3$~GeV
to be linearly dependent on $\sqrt{s}$ and causing
the shift of $M_Z$ of +0.17~MeV for lepton pair production 
and -0.17~MeV for hadronic final states,
such that the net contribution of IFI to $M_Z$ 
is small due to the accidental cancellation.
Possible uncertainties in the $M_Z$ measurement due to IFI-like
missing \order{\alpha^2} and \order{\alpha\alpha_S} corrections
were discussed in ref.\cite{Jadach:1999gz} quantitatively,
concluding that the uncertainty of the IFI effect due to higher order corrections
is $\delta M_Z\simeq 0.1$~MeV and $\delta \Gamma_Z\simeq 0.1$~MeV
and of the total cross section at the top of the $Z$ peak
is merely $\delta\sigma/\sigma <10^{-4}$.
From the analysis of the IFI effect in ref.\cite{Jadach:1999gz}
it was also clear that the size of the IFI effect
and its uncertainty depend quite strongly 
on the centre of the mass (c.m.s) energy $\sqrt{s}$
and on experimental cut-offs, hence the role of the MC
in controlling it was already quite important in the LEP data analysis.
Similar analysis of the IFI effect in the lineshape observables
was performed in ref.~\cite{Bardin:1999gt}.
It was restricted to pure \order{\alpha^1} QED
semi-analytical calculations using 
\zfitter~\cite{Bardin:1999yd} and \topazz~\cite{Montagna:1998kp} programs, 
without soft photon resummation.

The mass of $Z$ boson comes from fitting of the cross section
across the $Z$ resonance (lineshape).
According to the final LEP1 and SLD data analysis of ref.\cite{ALEPH:2005ab},
the dominant QED contribution to the lineshape, 
which translates into $\delta M_Z\simeq 0.2$~MeV and $\delta \Gamma_Z\simeq 0.3$~MeV,
was not from photonic corrections, 
but rather from the light lepton and quark contributions calculated
in ref.\cite{Arbuzov:2002rp}.
In the FCC-ee context,
this contribution should be cross-checked and carefully re-analyzed.

\subsection{Luminosity measurement}

The overall error of the luminosity measurement
quoted by all LEP experiments for LEP1 was below 0.05\%, 
with the best experimental luminosity error 0.034\% being that
of the OPAL collaboration.
The theoretical prediction of the QED dominated low angle Bhabha (LABH)
cross section in all LEP collaborations was based on the calculation from
\bhlumi~4.04 Monte Carlo event generator  published in ref~\cite{Jadach:1996is}.
Its precision, following ref.~\cite{Ward:1998ht},
was quoted to be $\delta \sigma / \sigma \simeq 0.061\%$.
In fact, the luminosity cross section was the only observable in LEP experiments
for which the theory uncertainty was bigger 
than the experimental statistical and systematic error.
As seen in tab.~2 of ref.~\cite{Jadach:2018jjo} its biggest component
was in fact not of the QED origin but due to the vacuum polarisation effect,
which was calculated using low energy experimental hadronic data.
Since in the recent years the error of vacuum polarisation contribution
was reduced significantly, the updated error of \bhlumi~4.04 predictions
is now estimated to be 0.038\%
(see the same Table in ~ref.~\cite{Jadach:2018jjo})
and is now dominated by uncertainties due to missing
higher order perturbative QED corrections.

The principal luminosity measurement at FCC-ee will be done using
the same type of the low angle Bhabha process as in LEP experiments,
using a similar calorimetric detector~\cite{MogensDam:2018}.
Prospects of improving QED predictions for the FCC-ee luminometry
will be discussed in the following.

In ref.~\cite{Jadach:2018jjo}) (Table 3) it was pointed out
that due to twice wider angle of the FCC-ee Bhabha luminometer (64--86 mrads)
than at LEP, the actual theoretical error according
to the present state of the art (inherited from LEP) would be in fact 0.090\%
due to the bigger $Z$-exchange contribution.
However, this error can be reduced to the negligible level already now
using the \bhwide\ Monte Carlo~\cite{Jadach:1995nk},
see ref.~\cite{Jadach:2018jjo}) for the details.

\subsection{EW mixing angle from charge and spin asymmetries at LEP}

All charge and spin asymmetries measured at LEP have their errors
dominated by the statistical error%
\cite{Abbaneo:2001ix,ALEPH:2005ab,Alcaraz:2006mx,Alcaraz:2007ri,Schael:2013ita}.
Let us characterize briefly
these (yet subdominant) QED corrections in the LEP measurements.

Charge and spin asymmetries depend on the ratios of the $Z$ couplings.
With the usual simplifications~\cite{Abbaneo:2001ix} and/or
introduction of pseudo-observables (PO)\cite{Bardin:1999gt},
in which QED effects are ``deconvoluted'',
it is convenient to express all charge and spin asymmetries
(following notation of ref.~\cite{ALEPH:2005ab})
in terms of
\begin{equation}
  {\cal A}_f = \frac{ 2g_{Vf}/g_{Af} }{1 +(g_{Vf}/g_{Af})^2},\quad
  \frac{g_{Vf}}{g_{Af}} = 1-4 |Q_f| \sin^2 \theta^f_{eff}.
\end{equation}
In this work we consider leptonic charge asymmetry and two $\tau$ spin asymmetries
\begin{equation}
   A^l_{FB}                      = \frac{3}{4} {\cal A}_e {\cal A}_l,\quad
   A_{\rm FB}^{pol.\tau}           = -\frac{3}{4} {\cal A}_e,\quad
   \langle {\cal P}_\tau \rangle = -{\cal A}_\tau .
\end{equation}
The above defines convenient pseudo-observables (pseudo-parameters)
$\sin^2 \theta^l_{eff}, l=e,\mu,\tau$
which allow us to compare information on $Z$ couplings from various asymmetries.
Using $\sin^2 \theta^l_{eff}=0.2315$ of tab.~\ref{tab:table1}
($ A_l = 0.1472$) and assuming lepton universality,
the following simple relations relate uncertainties of asymmetries
and of the EW mixing angle:
\begin{equation}
\delta\sin^2 \theta^l_{eff} = 0.5692\cdot \delta A^l_{FB},\;
\delta\sin^2 \theta^e_{eff} = 0.1676\cdot \delta A_{\rm FB}^{pol.\tau},\;
\delta\sin^2 \theta^\tau_{eff} = 0.1257\cdot \delta \langle {\cal P}_\tau 
\rangle.
\label{eq:sW2trans}
\end{equation}

How big were estimates of the theory uncertainties
$\delta A^l_{FB}$, $\delta A_{\rm FB}^{pol.\tau}$ 
and $\delta \langle {\cal P}_\tau \rangle$  due to QED effects
in data analysis at LEP near the $Z$ resonance and above?

The first systematic study of the QED uncertainties in $\afb$ 
was attempted in the initial LEP workshop~\cite{Bohm:1989pb},
in particular the value of the strongly suppressed IFI contribution
$A^{\mu, IFI}_{FB}= 5\cdot 10^{-4}$  (following ref.~\cite{Jadach:1988zp})
was established.

As already said, at the $Z$ peak ref.~\cite{ALEPH:2005ab} cites
the difference $\delta A^l_{FB}=10^{-4}$ between \zfitter\ and \topazz\
as a theory uncertainty estimate.
A closer look into ref.~\cite{Bardin:1999gt} on which ref.~\cite{ALEPH:2005ab}
is based reveals that the story is more complicated.
For IFI switched off,
from the differences \zfitter$-$\topazz\ and due to the change of the ISR
radiation function (from the factorized to additive form) shown in 
ref.~\cite{Bardin:1999gt}
one may indeed quote $\delta A^l_{FB}(M_Z)\simeq 10^{-4}$ and 
$\delta A^l_{FB}(M_Z\pm 3GeV)\simeq 3\cdot 10^{-4}$, especially for loose cut-offs%
\footnote{ Similar earlier study of ref~\cite{Was:1989ce} based on
  comparisons of analytical calculations versus of \koralz\ Monte Carlo was quoting
  $\delta A^l_{FB}(M_Z)=0.0005$ and $\delta A^l_{FB}(M_Z\pm 3GeV)=0.005$.
}.

On the other hand, for IFI switched on,
the difference between \zfitter\ and \topazz\ in ref.~\cite{Bardin:1999gt}
for realistic cut-offs, within $|\sqrt{s}-M_Z|<3GeV$, were quite sizable,
of order of a few per mille%
\footnote{The authors of ref.~\cite{Bardin:1999gt} were recommending subtraction
   of IFI using MC programs like KORALZ, with complete \order{\alpha^1}.
}.
Fortunately, in ref.~\cite{Christova:1999cc} the source of  these discrepancies 
between IFI implementation  in \zfitter\ and \topazz\ were identified and
new precision estimates
$\delta A^l_{FB}(M_Z)=0.0002$ and  $\delta A^l_{FB}(M_Z\pm 3GeV)=0.0013$
were provided.
However, this estimate is not reliable for higher order effects in IFI,
because both \zfitter\ and \topazz\ implement essentially the same
additive combination of the ISR and FSR collinear radiator functions 
with \order{\alpha^1} results, integrated analytically over a single real photon
within some simple cut-offs.
Hence the above estimates really represent the technical precision 
of both programs and not their physical precision.

As already noted in ref.~\cite{Bardin:1999gt} the necessary next step should be 
simultaneous exponentiation of ISR and IFI.
This goal was achieved, almost in parallel with the above works,
in the \kkmc\ program\cite{Jadach:1999vf}.
In ref.~\cite{Jadach:2000ir} the comparison of \kkmc\ with \zfitter\
and \order{\alpha^1} \koralz\ has shown that indeed for loose cut-offs
one may conservatively estimate higher order QED effects 
to be $\delta A^\mu_{FB}(M_Z)=0.0005$.
According to eq.~(\ref{eq:sW2trans})
this estimate translates into $\delta\sin^2 \theta^e_{eff}= 2.8\cdot 10^{-4}$.
This QED uncertainty was about factor two below the LEP experimental 
precision at the $Z$ peak, see Table~\ref{tab:table1}.

Later on, at the LEP2 energies near and above the $WW$ threshold the experimental
LEP precision of $\afb$ was worse due to smaller statistics, 
however, the IFI contribution was again quite
important because it was not suppressed.
For mild experimental cutoffs it could reach a few percent.
The precision of the traditional \order{\alpha^1} calculations of IFI
was believed to be $\sim 1\%$, see ref.~\cite{Christova:1999cc}.
More systematic studies of the QED uncertainty, including IFI, at LEP2 energies
were done in refs.~\cite{Kobel:2000aw,Jadach:2000ir,Schael:2013ita}.
In ref.~\cite{Kobel:2000aw} the overall QED uncertainty based mainly
on comparisons of  \kkmc\ with \zfitter\ and \koralz\ was estimated to be
$\delta A^\mu_{FB}(M_Z)=0.004-0.005$ and for the IFI component 
$\delta A^{\mu, IFI}_{FB}(M_Z)<0.003$.
The more detailed study of ref.~\cite{Jadach:2000ir} has
concluded that the overall QED uncertainty (including IFI) 
of the charge asymmetry  prediction from \kkmc\ 
is $\delta A^\mu_{FB}(M_Z)=0.002$ 
over the entire LEP2 energy range.

\subsection{The invisible $Z$ decay width from cross section 
            and radiative return in LEP experiments}
\label{sec:Zinv}

In LEP experiments the measurement of the $Z$ invisible decay width was quantified
as the deviation from the Standard Model expectation 
of the neutrino number and family generation number, $N_\nu=3$.
This parameter was measured to be $N_\nu=2.984\pm 0.008$ \cite{ALEPH:2005ab}.
It was deduced mainly from the total cross section at the $Z$ peak $\sigma(M_Z)$.
The dominant contribution to its uncertainty was the luminosity error.
According to ref.~\cite{ALEPH:2005ab} the luminosity error contributed
$\delta N_\nu = 7.5\; \frac{\delta {\cal L}}{{\cal L}}$.
For $\frac{\delta {\cal L}}{{\cal L}}= 6.1\cdot 10^{-4}$
it gives $\delta N_\nu = 4.6 \cdot 10^{-3}$.
At the FCC-ee the above luminosity uncertainty will be again the dominant one.
Getting $\delta N_\nu = 1\cdot 10^{-3}$ will require
the luminosity error to be improved by a factor $\sim 5$,
down to $\frac{\delta {\cal L}}{{\cal L}}= 1\cdot 10^{-4}$,
which looks feasible, see Sect.~\ref{sec:Lumi2}.

Another kind of measurement, the so called $Z$ radiative return (ZRR)
was also exploited in LEP experiments,
see for instance ref.\cite{Abbiendi:2000hh},
where the less precise result $N_\nu=2.69\pm0.15$ was quoted.
In the ZRR process one is examining the energy distribution of a photon
emitted above some minimum angle from the beams,
at the c.m.s. energy well above the $Z$ resonance.
Such a photon is also seen for $Z$ decaying into neutrinos or other
``invisible'' particles.
The $Z$ resonance peak is seen in the energy distribution of the ZRR photon
and  $N_\nu$ is obtained by means of comparing the integrated cross section 
of this process with the result of a reference Monte Carlo program.
The error of the above LEP measurement  $\delta N_\nu = 0.15$ was dominated
by the statistical error.
The QED component of this error was $\delta N_\nu = 0.06$,
according to $\delta N_\nu = 3\delta \sigma/\sigma$,
where $\delta \sigma/\sigma = 2\%$ was the QED/SM error
attributed to the cross section calculations
of \koralz\cite{Jadach:1993yv,Colas:1990ef}
and {\tt NUNUGPV98}\cite{Montagna:1998ce} programs.
Such a precision was then quite satisfactory in view of the large statistical error.
The calculation of ZRR using the more advanced \kkmc\ program\cite{Jadach:1999vf}
was not yet available at that time.
Another theoretical study was done in ref.~\cite{Igarashi:1986ht}.
The improvement at FCC-ee down to $\delta N_\nu = 1\cdot 10^{-3}$ would require 
$\delta \sigma/\sigma = 3\cdot 10^{-4} $ precision 
for the MC programs calculating ZRR process (factor $\sim 60$ improvement).
However, with the advent of  \kkmc\  this factor is clearly an overestimate,
see the following discussion.

With FCC-ee precision the error of the
luminosity measurement will enter into the game
and will have to be improved as well.
On the other hand, thanks to high luminosity of FCC-ee
it would be possible to eliminate the dependence on the luminosity error
in the ZRR method by means of using the ratio of the photon distribution
with invisible $Z$ decay and with $Z$ decaying in the muon pair.
Such a method could not be exploited at LEP due to the limited
statistics of the ZRR with muon pairs.
At FCC-ee it makes sense and the prospects of its precision in terms of $N_\nu$ 
are discussed in the following Sect.~\ref{sec:Zinv2}.

\subsection{QED in $W$-pair production at LEP, $M_W$ measurement}
\label{ssect:WWatLEP}

The role of QED in the LEP2 data analysis near and above the $WW$ threshold
was somewhat different than near $Z$ pole
-- there was no systematic attempt to ``deconvolute'' universal 
(process independent) QED effects
and develop the technique of pseudo-observables --
QED was usually kept together with pure electroweak corrections
and effects due to semi-classical QED interaction of 
two massive $W$ near the production threshold, 
the so-called Coulomb effect, had to be included in the game.
Generally, there are four classes of QED effects in $e^+e^-\to W^+W^-$
or $e^+e^-\to 4f$ process: 
initial state corrections (ISR), 
final state corrections (FSR) in the decays of two $W^\pm$,
final state Coulomb corrections (FSC) and
the so-called non-factorisable interferences%
\footnote{They factorise at the amplitude level,
 but become cumbersome after squaring amplitudes.}
between ISR and FSR in two $W^\pm$ decays (IFF).
The IFF corrections are suppressed due to relatively long lifetime of $W$'s.
The effects due to ISR are numerically the biggest but easier to control,
while the FSR effects can be also quite sizable for typical experimental cut-offs.

The experimental error of the measurement of the $W$ boson mass 
from the total cross section of $e^+e^-\to W^+W^-$ near threshold 
at LEP2 experiments~\cite{Schael:2013ita}
was $\delta M_W=200$~MeV due to poor statistics of the data,
quoting theoretical error in this measurement as negligible.

Much better experimental precision of $\delta M_W=34$~MeV
was achieved in LEP2 experiments~\cite{Schael:2013ita} from $W$ mass reconstruction 
for $l\nu q\bar{q}$ and $q\bar{q}q\bar{q}$ final states.
For this method, the uncertainty of the \order{\alpha^1} theoretical calculations
used for $e^+e^-\to W^+W^-$ process was estimated in ref.~\cite{Schael:2013ita}
from the difference between \kandy~\cite{Jadach:2001mp,Jadach:2000kw}
and \racoon~\cite{Denner:2000bj,Denner:2002cg} programs.
In this way, for $l\nu q\bar{q}$ channel $\delta M_W=8$~MeV was obtained
and for $q\bar{q}q\bar{q}$ it was $\delta M_W=5$~MeV.
In particular, the uncertainty due to ISR radiation alone $\delta M_W=1$~MeV
was obtained in ref.~\cite{Schael:2013ita}
using \kandy, \racoon\ and \wphact~\cite{Accomando:2002sz}
by means of switching on/off \order{\alpha^3L_e^3} ISR contribution.

The general discussion of the theory issues in the precision SM
calculations for the $e^+e^-\to W^+W^-$ process can
be found in ref.~\cite{Grunewald:2000ju}.
However, there were many other works focusing on various specific issues.
For instance the detailed analysis of the ISR effects
can be found in ref.~\cite{Skrzypek:1995ur}.
The ISR effect on the total cross section of $e^+e^-\to W^+W^-$ near the threshold,
at 160~GeV is $-$28\% and merely $-$7.5\% at 205~GeV.
However its uncertainty deduced from switching on/off non-IR \order{L^3_e \alpha^3}
is not bigger than of $2.5\cdot 10^{-4}$ at 160~GeV and $2.1\cdot 10^{-4}$ at 205~GeV.
The entire $-$28\%  ISR effect at 160GeV would translate into huge 
$\delta M_W \sim 400$~MeV, 
while its uncertainty is worth only $\delta M_W \sim 0.2$~MeV.

Dedicated analysis of the QED and non-QED effects in the $M_W$ reconstruction
from the final states can be found in ref.~\cite{Jadach:2001cz}
\footnote{A short overview of the experimental methodology 
  is also included there.}.
The uncertainty due to not included higher order 
and nonleading ISR effects was estimated 
in Table 1 of ref.~\cite{Jadach:2001cz} to be $\delta M_W < 1$MeV,
the FSR uncertainty was rated as $\delta M_W \sim 2$MeV
and the IFF (following ref.~\cite{Chapovsky:1999kv}) at $\delta M_W < 2$MeV.
The above analysis is based on the results from \kandy\ toolbox 
of the Monte Carlo and semianalytical programs \cite{Jadach:2001mp}.

SM calculations for $e^+e^-\to 4f$ process actually
used in the LEP2 data analysis were all in form
of the Monte Carlo event generators and they included both QED
and the remaining EW corrections.
In the following we shall provide brief descriptions of them
commenting also on how the QED part was organized/implemented.
There were two such MC codes with ${\cal O}(\alpha)$ corrections 
to the signal doubly resonant CC03 graphs calculated respectively 
in the leading pole (LPA) or double pole (DPA) approximation%
and with the tree level matrix element for the remaining
$e^+e^-\to 4f$ contributions:
\begin{itemize}
\item 
\yfsww3 v.1.16 \cite{Jadach:1996hi,Jadach:2001uu} generates the signal process
$e^+e^-\to W^+W^-\to 4f$ according to the LPA scheme,
exploiting the \order{\alpha^1} calculations of
Refs.~\cite{Fleischer:1988kj, Kolodziej:1991pk,Fleischer:1991nw, Fleischer:1994sq}
and with $W^{\pm}$ decays simulated independently.
Multiphotonic radiation for production part, $e^+e^-\to W^+W^-$, 
is implemented in the YFS framework (EEX scheme).
The hard photon ISR is corrected to the \order{L_e^3 \alpha^3}.
The FSR in two separate $W$-decays is handled by \photos.
The \koralw\ 1.42 \cite{Jadach:1998gi} code,  
which simulates the complete $e^+e^-\to 4f$ process at the tree level,
has been combined on the event per event basis with the \yfsww3 code. 
This way a concurrent MC called \kandy\ of ref.~\cite{Jadach:2001mp} has emerged,
which simulates both the complete four-fermion final states
and includes ${\cal O}(\alpha)$ corrections to the W-pair production and decay.
The semianalytical program \korwan\ for the improved Born approximation (IBA)
is included for testing the main MC.
\item
\racoon\ \cite{Denner:2002cg}
simulates the complete $e^+e^-\to 4f$ Born-level process,
and the single real photon emission process $e^+e^-\to 4f+\gamma $ 
and  implements $\cal{O}(\alpha)$ electroweak
virtual corrections in the DPA scheme~\cite{Denner:2005es, Denner:2005fg}
exploiting one-loop calculations for on-shell $WW$  production and decay. 
The ISR radiation is based on the QED collinear structure 
functions to second order with soft photon exponentiation.
It includes also semianalytical program for
the Improved Born Approximation (IBA)~\cite{Denner:2001zp},
based on CC03 graphs (doubly resonant) 
and universal corrections to these graphs, i.e.
the Coulomb correction at the threshold, 
the running of effective couplings and collinear ISR.
\end{itemize}
Both programs include Coulomb corrections for off-shell $W^\pm$ bosons.
According to expert comparisons of the above two aproaches
in refs.~\cite{Grunewald:2000ju,Placzek:2002ft},
DPA and LPA methods%
\footnote{Differences between two variants of LPA are explained in ref.~\cite{Jadach:2000kw}.}
of inserting \order{\alpha^1} EW corrections
into doubly-resonant part of the matrix element in the $e^+e^-\to W^+W^-$ process
are basicaly equvalent.
The main differences are in the implementation of the QED part of the matrix element.
The complete ${\cal O}(\alpha)$ corrections to $e^+e^-\to W^+W^-$ cannot differ --
it was checked in ref.~\cite{Grunewald:2000ju} that
the difference  for the total cross section
between implementations in \yfsww3 and \racoon\ 
of the virtual plus soft corrections 
in two programs is below 0.01\%.

The overall agreement of \yfsww3 and \racoon, 
including all physical effects,
is of the  order of 0.3\% for the total cross section at 
200~GeV~\cite{Grunewald:2000ju}.
The differences of the \order{\alpha\Gamma_W/M_W} are expected
due to inefficiency of DPA/LPA.
However, most of these differences arise
from the different treatment of the QED corrections in both approaches,
so it is interesting to look into them.

QED calculations in \racoon\ rely on the massless 
$e^+e^-\to 4f+\gamma$ matrix element,
hence the complete standard full phase space cannot be used.
Instead, two methods are used to deal 
with emission of photons collinear with fermions.
In one of them the phase space is ``sectorized'',
i.e. the real photon phase space is restricted 
with minimum angle of the photon to fermions
(also minimum photon energy to exclude IR divergence)
and the contribution below the minimum angle is integrated over analytically,
recovering correct fermion mass dependence.
In another method~\cite{Denner:1999kn} a QED variant
of the Catani-Seymour (CS) subtraction scheme~\cite{Catani:1996vz} is used.
In both these methods collinear ISR or FSR photon angular distributions 
are integrated over and effective longitudinal momentum distributions 
(effective radiator functions and/or PDFs) arise.
Their original \order{\alpha^1} version is upgraded to include LO+NLO
corrections up to \order{L_f^3\alpha^3}.
Soft photon resummation is also included in the effective radiator functions,
but multiphotons are not present explicitly in the MC events.
Non-factorisable interferences between the production and two decays of $W$s
are reproduced up to \order{\alpha^1} 
(in the soft photon approximation)

The methodology of QED treatment in \yfsww3 or \koralw\ is quite different
from that in \racoon.
Multiple photons are present explicitly in the MC events.
The matrix element is constructed following EEX variant of the YFS soft photon
factorisation and resummation.
Non-soft higher order LO and NLO collinear photon universal contributions
are added to matrix element in the exclusive (unintegrated) form,
without collapsing to collinear photon distributions, to a $\delta(\theta)$ function.
The multiple photon emission is included in \yfsww3 for ISR and FSR out of $W$s,
while in \koralw\ it is restricted to ISR.
Single and double photon emissions in the decays of $W$s are added 
using \photos\cite{Jadach:1993hs,Barberio:1990ms} program%
\footnote{
  Later on in ref.~\cite{Placzek:2003zg}, multiphoton radiation based on the YFS 
  for single $W$ decay was implemented in another MC code
  \winhac~\cite{Placzek:2003zg} developed  for LHC.
  This could be easily adapted to the \yfsww3 code.}.
Non-factorisable interferences between production and two decays of $W$s are not included.
The YFS technique of factorizing and resuming the universal
QED corrections employed in \yfsww3 provides for a clear separation of the  universal
QED corrections from the rest of the SM radiative corrections at any order,
hence it is a very good candidate for the future inclusions of the 
non-QED EW corrections beyond the \order{\alpha^1}, or in any attempt
of better organisation (deconvolution) of the existing \order{\alpha^1} calculations.

After LEP2 data were analyzed,
the complete ${\cal O}(\alpha)$ corrections to $e^+e^-\to 4f$ have been also completed
in Refs.~\cite{Denner:2005es, Denner:2005fg} for the 
final states without repeated flavours and without CKM-suppressed states
\footnote{%
  There is no MC event generator implementing
  the complete ${\cal O}(\alpha)$ correction to the $e^+e^-\to 4f$ process.
}.
The comparison of the above new calculation with older LPA/DPA results
gives new insight into their uncertainty beyond universal QED corrections.
Generally these newer calculations confirm the precision estimates of the LPA/DPA
approaches of the LEP2 era.
Near the $WW$ threshold the difference between the DPA/LPA 
and the complete 4-fermion ${\cal O}(\alpha)$ is $\sim$ 2\% of the $4f$-Born.
Far from the $WW$ threshold the difference DPA vs.\  
complete 4-fermion ${\cal O}(\alpha)$ 
is smaller, drops to below 0.5\%, in accordance 
with the stated precision of the relevant MC codes.
(It rises back to 1--2\% at 1--2 TeV energies.)

Refs.~\cite{Denner:2005es, Denner:2005fg} provide also 
estimation of the missing higher order EW corrections,
with the important conclusion that higher order EW corrections
are dominated by the QED contribution $\alpha^2 \log(m_e^2/s)$
and are estimated at $\leq 0.1\%$ for energies below 500 GeV.
The other interesting finding is
that the contribution of the higher order Coulomb effect is of the order of 
$0.2\%$ at the threshold.%
\footnote{
  The QCD effects must be also included: ${\cal O}(\alpha_S)$ corrections,
  uncertainties due to matching with parton shower, Bose-Einsein and colour
  reconnections.
}.
The total precision of these four-fermion ${\cal O}(\alpha)$ calculations has
been estimated by the authors to be a few per mille.

Another calculation specialised to the near $WW$ threshold energy,
using the effective field theory (EFT) technique 
is reported in Refs.~\cite{Beneke:2007zg,Actis:2008rb}.
The dominant NNLO corrections to four-fermion process 
($\mu^-\bar\nu_\mu u \bar{d}$) 
were calculated using EFT for unstable particles --
the best calculation was nick-named as N$^{3/2}$LO$^{\rm EFT}$,
because in EFT a different expansion parameter 
(relativistic velocity of $W$ in the $WW$ rest frame)
is used to count the strength of particular corrections.
The drawback of the EFT method is that it provides the {\em inclusive} results only.
The effect of these pure N$^{3/2}$LO$^{\rm EFT}$ corrections 
on the W mass is estimated as 3 MeV. 

The complete set of graphs for the $e^+e^-\to 4f$ process at the tree level
was implemented in several MC codes~\cite{Grunewald:2000ju}.
Two of them were used in the actual data analysis of LEP2:
standalone version of \koralw\ \cite{Jadach:1998gi}
and \wphact~\cite{Accomando:2002sz}.
The latter one implements the so-called
''Fermion Loop'' gauge restoring scheme of ref.~\cite{Beenakker:1996kn}.

The present state of the art (mostly inherited from LEP)
can be summarized as 0.2\% theoretical precision for
the total cross section of the $W$-pair production process
in the entire energy range.
If statistics was not limited at LEP2,
that would translate into $\delta M_W \simeq 3$~MeV 
for the $M_W$ measurement from the threshold cross section.
Looking at the future needs of the FCC-ee we can see that a factor of 10 improvement 
in precision is needed relative to the present state of the art. This is a 
relatively moderate goal as compared to other observables, see 
Table~\ref{tab:table1}. We will discuss its feasibility in 
Sect.~\ref{ssect:WWatFCC}.

\newpage
\section{Prospects of the QED calculation improvements in the FCC-ee measurements}

In contrast to the previous section, where we have elaborated on the present
state of the art in the calculations of QED effects for observables selected
in Table~\ref{tab:table1},
in the following we shall list all possible developments
in QED calculation needed for analysing future data at FCC-ee.
Obviously, this will be to some extent speculative and it is quite
probably that we are going to miss some
new calculation fronts or methods which will really emerge in the future.

\subsection{Cross section near the $Z$ resonance and mass of Z}

As seen in tab.~\ref{tab:table1}, the hadronic cross section 
$\sigma^0_{\rm had}$ at the $Z$ peak was measured with 0.09\% total error
with 0.06\% component dominated by the theoretical uncertainty in the luminosity measurement.
Anticipated factor 6 improvement at FCC-ee is expected
due to reduction of luminosity error down to 0.01\%,
both in theory and experiment.
According to ref.~\cite{Jadach:1999pp} the uncertainty due to QED ISR
uncertainty in fitting $\sigma^0_{\rm had}$ to the experimental lineshape is $0.02\%$,
while the uncertainty of the IFI contribution~\cite{Jadach:1999gz} is even smaller.
The obtained uncertainty was very conservatively.
For the ISR photonic corrections
it was obtained by means of switching on/off parts of the \order{L_e^3\alpha^3}
and \order{L_e\alpha^2} components in the effective 
radiator function and by taking half of them.
One should better evaluate analyticaly known corrections of \order{L_e^4\alpha^4}
\cite{Przybycien:1992qe} and \order{L_e^0\alpha^2} \cite{Berends:1987ab}
and estimate unknown corrections of \order{L_e^2\alpha^3}.
Most likely they are at the level of $10^{-5}$.
According to ref.~\cite{Arbuzov:2002rp} 
quasi-photonic light fermion contributions 
(from electron, muon, $\tau$ pairs and light quarks) 
to the ISR radiator function also contribute
in addition $0.02\%$ to the uncertainty of $\sigma^0_{\rm had}$.
Most likely this is an overestimate because this uncertainty
is approaching half of the effect itself.

The optimistic point of view might be that for
the hadronic cross section $\sigma^0_{\rm had}$, 
due to almost 100\% detector acceptance,
known QED calculations are sufficient to reach the 0.01\% precision
needed for FCC-ee -- only better testing of the existing calculations
and better estimates of the missing h.o. corrections are needed.
In particular mixed corrections QED-QCD should be reexamined.
The improvements on light fermion pair corrections require some
effort but this looks feasible.
However, in order to get back to the LEP situation where
QED uncertainties were factor $\sim 2-3$ smaller than experimental
one then more work would be needed.
This would also imply improvements in the luminosity measurement
beyond what is described in Sect.~\ref{sec:Lumi2},
which will be hard to achieve.

On the other hand, $R^Z_l=\Gamma_h/\Gamma_l$ is free of
the luminosity problem but more sensitive to QED uncertainties
due to relatively complicated acceptance, 
with the isolation cones around beams, 
more restrictive cut on the total photon energy (or on acollinearity),
and the $t$-channel contribution for the $e^+e^-$ final state.
The factor 12 improvement in tab.~\ref{tab:table1} for  $R^Z_l$
from 0.60\% down to 0.05\%, or even more if the QED uncertainty is to be kept
below the experimental error will require better calculations.
The most demanding will be the FSR class of corrections, but ISR and IFI
will have to be studied at the level close to the 0.01\% precision level.
Semianalytic calculations of the class like 
\zfitter~\cite{Bardin:1999yd} and \topazz~\cite{Montagna:1998kp}
used at LEP will not be sufficient for the task.
The upgraded version of \kkmc\ or a similarly accurate Monte Carlo program
will have to be developed and used.
Dedicated study is needed to determine what level of the improvement
of the perturbative calculation is really needed to match FCC-ee precision of $R^Z_l$.
Most likely it will be one order more accross precision boundaries in Fig.~2
at various perturbative orders in powers of $\alpha$ and mass logarithms
for non-soft (IR-finite) QED corrections
in the matrix element of the MC program.

The mass of the $Z$ comes from fitting the cross section
across the $Z$ resonance (lineshape) such that most of the QED effects are removed.
It is the hadronic cross section which matters mostly.
In the final LEP1 and SLD analysis of ref.\cite{ALEPH:2005ab}
the estimation of QED uncertainties in the measurement of $M_Z$
was taken from refs.~\cite{Jadach:1999pp} for ISR,
from \cite{Jadach:1999gz} for IFI and from \cite{Arbuzov:2002rp}
for light fermion pairs.
According to ref.~\cite{Jadach:1999pp} the uncertainty due to
photonic ISR corrections to light fermion pairs process
is negligible because it is weakly dependent on the c.m.s. energy.
This statement has to be cross-checked.
The effect of IFI according to ref.~\cite{Jadach:1999gz}
is sizable ($ \sim 0.17$~MeV) and its uncertainty induces
$\delta M_Z, \delta \Gamma_Z\leq 0.1$MeV errors.
Light fermion pairs also contribute $\delta M_Z, \delta \Gamma_Z\leq 0.1$MeV.
Both of these QED uncertainties are therefore of the size of the FCC-ee experimental errors,
hence for the comfortable data analysis they should be reduced by at least a factor 2-3.
The IFI effect in the $M_Z$ measurement has a similar size
and the opposite sign for the leptonic and hadronic final states.
It is strongly dependent on the experimental cut-offs.
Calculations of \order{\alpha^1} without exponentiation
used at LEP for evaluating IFI will not be sufficient at the FCC-ee precision.
For extracting $M_Z$ and $\Gamma_Z$ from $\sigma_h(s)$ near the $Z$ peak
further progress in reducing uncertainty of IFI
and light fermion pair contributions in $M_Z$ beyond the LEP state of the art
will be necessary.
It is possible that for this particular purpose
hybrid approach of LEP era combining the use
of MC event generators and semianalytical programs like ZFITTER/TOPAZ0 will still work.
However, the alternative approach based entirely on the MC event generators
will serve as crosscheck and it will also have an advantage to work for
more difficult case of charge and spin asymmetries.
For the moment \kkmc\ is the leading candidate for further studies of the IFI effects.
However, its CEEX matrix element should be upgraded to include 
\order{\alpha^3L_e^3} and non-IR parts of \order{\alpha^2L_e^1} pentaboxes.
For improvements of the light fermion pair contributions one should exploit programs
dedicated to four fermion final states, the same as for the production $WW$ process.

Summarizing, reduction of QED uncertainties in $\sigma^0_{\rm had}$ below 0.1\%,
in $R^Z_l$ below 0.05\%, in $M_Z$ and $\Gamma_Z$ below 0.1~MeV is definitely feasible,
but requires more work and improvements of the existing MC and semianalytical programs.
Improvements on light fermion pair contributions seems to be the most urgent.

\subsection{Charge and spin asymmetries at FCC-ee}

Let us concentrate on symmetries at the $Z$ peak, 
where they will be measured most precisely.
As seen in Table~\ref{tab:table1}, the charge and spin symmetries will be measured
up to a factor $\sim 50$ more precisely than at LEP, which will require MC tools
for calculating differential distributions with realistic experimental cut-offs
including \order{\alpha^2}, possibly even \order{\alpha^3},
electroweak, QCD and non-universal QED corrections,
while resummed QED universal corrections will have to be
included at even higher orders.
Concerning feasibility of such higher order  QED and EW calculations, 
let us cite statement in the summary 
of the recent workshop proceeding~\cite{Blondel:2018mad}:
``The techniques for higher order SM corrections are basically understood, 
but not easily worked out or extended ...  
We anticipate that at the beginning of the FCC-ee campaign of precision measurements, 
the theory will be precise enough not to limit their physics interpretation.''

Presently, the only MC tool (event generator) which provides 
predictions for all charge and spin asymmetries for arbitrary 
experimental cut-offs, including \order{\alpha^1} EW corrections,
complete \order{\alpha^2} QED corrections 
and soft-resummed universal QED corrections to infinite order,
is the \kkmc\ Monte Carlo~\cite{Jadach:1999vf},
at the $Z$ peak and far away from it.
Its precision goes far beyond what was needed at LEP,
as seen for example in the recent study 
of the IFI contribution to $A_{FB}(M_Z\pm  3.5{\rm GeV})$,
where the precision $\delta A_{FB} \le 10^{-4}$ 
was verified using auxiliary calculations.
%
However, for the FCC-ee experimental precision of asymmetries
quoted in Table~\ref{tab:table1}, the QED part of matrix element
has to be upgraded to include the next orders,
up to \order{\alpha^3L_f^3}
in the CEEX matrix element,
and EW corrections should be upgraded in the MC matrix element
to the level of known complete \order{\alpha^2} 
corrections~\cite{Freitas:2014hra,Freitas:2014owa,Dubovyk:2018rlg,Dubovyk:2019szj}.
Needless to say, \tauola\ MC used in \kkmc\ for 
$\tau$ lepton decays will require
additional testing and development.
Of course, development from the scratch
of another MC program of similar quality as \kkmc\
would be of great help in the solid independent validation of the required precision.

On the other hand, let us point out to some important 
problems with the model-independent
representation of the data in a form of simple pseudo-observables
like in Table~\ref{tab:table1},
where the information on the fermion couplings to $Z$ 
extracted from charge and spin asymmetries
is quantified in terms of a single parameter, $\sin^2\theta_W^{eff}$.
At LEP it was possible.
The big question is whether for much higher precision at FCC-ee
it will be still possible?
In the methodology of ref.~\cite{ALEPH:2005ab} a parameter
$\sin^2\theta_W^{eff}$ is just in one-to-one correspondence with the ratio
of the (real) vector and axial $Zf\bar{f}$ couplings in the effective Born,
which was fit to the $e^+e^-\to  f\bar{f}$ data
(taking into account factorisable QED corrections).%
\footnote{According to terminology of subsection \ref{ssect:ewpos}
  it is example of a ''EW pseudo-parameter'', EWPP.}
Whether this kind of ''effective Born'' will be still effective in parametrising FCC-ee data
in the (SM-)independent way is an open question.
Some known SM effects, could invalidate it, if they are not numerically small 
as compared to FCC-ee data precision.
Most important among them is the $s$-channel non-resonant contribution,
which at the $\sqrt{s}=M_Z$ drops out because it is almost exactly real,
while the $Z$ contribution is purely imaginary.
The reduction of the effective c.m.s. energy due to ISR would possibly invalidate that.
The electroweak $WW$ and $\gamma Z$ boxes also go 
beyond the simplistic effective Born ansatz.
The $S$-matrix approach%
\footnote{See section C.2 in the recent report~\cite{Blondel:2018mad},
  summarizing on that.},
in which $Zf\bar{f}$ couplings form the residue of the $Z$-pole, 
provide nice justification of the effective Born ansatz.
These couplings in the $S$-matrix approach have small imaginary parts,
which are already included in the LEP definition of the effective Born
(see eq.~1.34 in ref.~\cite{ALEPH:2005ab}).
The effects due to EW boxes and imaginary parts of $Z$ couplings
were proven in ref.~\cite{Bardin:1999gt}
to be small as compared to LEP data precision, 
but this might be not true in FCC-ee environment.
Note also that they are unlikely to be affected by new physics.

In particular, although ISR is reducing c.m.s. energy by $\sim 100$MeV,
this effect is controlled to within 0.1~MeV \cite{Jadach:1999pp}, 
hence its effect in charge asymmetry
of order 1\% is probably controllable within the FCC-ee data precision.
The effect of QED initial-final state interference is suppressed%
\footnote{Outside the $Z$ resonance $\delta A^{IFI}_{FB} \simeq 0.01$.}
at the $Z$ peak, $\delta A^{IFI}_{FB}(M_Z) \simeq 0.0005$ \cite{Bohm:1989pb},
hence its calculation at the \order{\alpha^2}
is probably adequate to take it into account --
better quantitative study of the uncertainty of $A^{IFI}_{FB}$
beyond \order{\alpha^1} at the $Z$ peak is definitely needed.
Neglecting imaginary parts was found in ref.~\cite{Bardin:1999gt}
to induce $\delta A_{FB} \simeq 0.15\%$, 
hence it will be not negligible for
FCC-ee precision at the $\delta A_{FB} \simeq 10^{-5}$ level.
On the other hand, EW boxes are a little bit less problematic,
as they were found~\cite{Bardin:1999gt}
to induce a $\delta A_{FB} \simeq 10^{-4}$ effect only.

Summarizing, in the FCC-ee data analysis it is likely that the LEP-style
definition of $\sin^2\theta_W^{eff}$ will have to be either completely abandoned,
or replaced by some variant in which all the above effects are not neglected
but taken into account.
So far, there was no detailed analysis concerning this issue.
One may only guess that model-independent
subtraction of the  $s$-channel non-resonant contribution,
using data far away from the $Z$ resonance, may still work.
Fitting the imaginary parts of the $Z$ couplings in the effective Born
to data, that is treating them as additional pseudo-observables,
would be a brave decision.
Including $WW$ and $\gamma Z$ boxes and other \order{\alpha^1} EW (QCD) corrections
in the effective Born, i.e.\ removing them from data on the way to 
pseudo-observables
would really mean treating them the same way as QED 
and the major change in the meaning of the EW pseudo-observables
(better say the EW pseudo-parameters).
See also related discussion in subsection~\ref{ssect:ewpos}.

The $\tau$ lepton spin polarisation from the $\tau$ pair production at $Z$ is less prone
to the ISR QED effects, as it is weakly dependent on $\sqrt{s}$.
One specific source of the uncertainties for $\tau$ spin asymmetries is due
to the limited quality of the $\tau$ decays $\tau \to \nu_\tau \pi, \nu_\tau \rho$ 
used as a spin polarimeter.
The effects due to nonperturbative QCD and QED effects in the $\tau$ decays
should be controlled better than at LEP.
However, it looks that achieving $\delta\sin^2\theta_W^{eff} \simeq 5\cdot 10^{-5}$
using $\tau$ spin polarisation is within the reach of presently available
MC tools like \tauola\ and \photos\cite{Jadach:1993hs,Barberio:1990ms},
provided some extra testing is done~\cite{2019:cern-talk-was}.
In particular one should ''calibrate'' $\tau$ decay polarimetric features,
implemented in $\tau$ decay simulation MCs,
using high statistics $\tau$ decay samples from Belle experiments%
\footnote{At the c.m.s. energy of the Belle experiments single taus are not polarized,
  in the tau-pair production allows for this kind of testing.}.
The influence of EW boxes or imaginary parts in the effective $Z$ couplings
on the pseudo-observables related to the tau spin asymmetries was not studied
in the LEP era, because their precision was statistically limited.
More quantitative studies are needed.

Altogether, it is expected that $\sin^2\theta_W^{eff}$ from
the $\tau$ lepton spin polarisation will be measured at FCC-ee
as precisely as from charge asymmetries 
and will provide a powerful crosscheck on both measurements,
due to very different experimental systematics.
The QED induced uncertainties need to be re-examined both in the $\tau$ production
and decay processes, but no serious problems are expected.

Just one example of possible problem: 
the decay $\tau^\pm\to \pi^\pm\pi_0 \nu_{\tau}(\gamma)$ 
is an efficient $\tau$ spin polarimeter, 
where the additional photon emission has to be taken into account very precisely.
Hovewer, the above radiative process has to be distingushed from
the cascade decay the cascade decay 
$\tau^\pm \to \pi^\pm \omega \nu_\tau,\; \omega \to \pi^0 \gamma$,
which has a non-negligible combined
branching ratio of 0.0015~\cite{Tanabashi:2018oca}.
The future high precision Belle II data should be used to analyse precisely
this and other similar effects in the energy spectra of $\tau$ decay product
used for measuring $\tau$ polarization at the FCC-ee precision level.

Summarizing, it looks that in general the main problem is not so much in 
the better QED and SM calculations of various asymmetries,
but rather in the very survival of the methodology of pseudo-observables
(pseudo-parameters) used in the model-independent representation of data
for these asymmetries.

\subsection{Luminosity measurement}
\label{sec:Lumi2}

The luminosity measurement at FCC-ee will be based again on the low angle
Bhabha process~\cite{MogensDam:2018},
but one should also remember about the possible use of the $e^+e^-\to \gamma\gamma$
process for the FCC-ee luminometry, which is statistically limited,
but not vulnerable to uncertainty due to vacuum polarisation,
see Sect.~C.5 in ref.~\cite{Blondel:2018mad} for more details.

By the end of the LEP era any substantial improvement of the theoretical prediction
for the low angle Bhabha (LABH) process used to determine the LEP luminosity was 
effectively blocked by the large uncertainty of the vacuum polarisation, 
which was in fact of the QCD origin,
or more precisely due do experimental errors of the low energy hadronic data.
Since then, this uncertainty was reduced by a factor four and
by the time of FCC-ee experiments another factor two is probable.
With the present vacuum polarisation error, the LEP luminosity error would
reduce from 0.061\% down to 0.038\%, see tab.~2 in ref.~\cite{Jadach:2018jjo}.
In this way h.o. perturbative QED components in the uncertainty
of the LABH cross section get dominant.

However, any further progress will not be possible without solid control of
the so called {\em technical precision} i.e. any problems due to programming bugs,
mistakes in the MC algorithm, numerical instabilities.
The Monte Carlo event generator \bhlumi~4.04 of ref.~\cite{Jadach:1996is}
was the subject of many internal tests, in particular using elaborate comparisons
with semi-analytical calculations in ref.~\cite{Jadach:1996bx}.
These powerful crosschecks were unfortunately limited to not so realistic
kinematic cut-offs.
The FCC-ee luminometer detector similarly as at LEP will select/detect events
in a  way which cannot be dealt with using analytic methods.
The Monte Carlo is the only way to implement them in the theoretical calculations.
Due to all the above features of the luminosity detectors, the only reliable
way of controlling technical precision for realistic event selection is to
compare calculations from two different MC programs of the comparable quality.
In the LEP era there was no other MC program of the same quality as \bhlumi~4.04,
hence QED and technical uncertainties were lumped together.
Luckily, it seems that there are two such candidates for the next generation
improved MC's for LABH process, 
\bhlumi\ with the upgraded QED matrix element proposed in 
ref.~\cite{Jadach:1996is}
and another MC program \babayaga~\cite{Balossini:2006wc}, developed in recent years,
provided its QED matrix element is upgraded.

Assuming that the problems of the low precision of the vacuum polarisation 
and of poor control of the technical precision will be solved,
then yet another non-QED component in the budget of the expected theory uncertainty
of the LABH process following the LEP era state of the art starts to dominate.
As pointed out in ref.~\cite{Jadach:2018jjo}, due to the angular range
of the FCC-ee luminosity detector being factor two wider,
the contribution from the $Z$ exchange near the $Z$ peak will rise by factor of four.
If one estimates its uncertainty the same way as at LEP,
then the total luminosity error would jump to 0.097\%, see Table~3 in 
ref.~\cite{Jadach:2018jjo}.
However, this uncertainty is rather easy to reduce using the MC program
\bhwide\ of ref.~\cite{Jadach:1995nk} developed for the wide angle Bhabha process.
In ref.~\cite{Jadach:2018jjo} it was shown that the uncertainty due to the $Z$ exchange
can be reduced substantially, perhaps down to 0.001\%.

In this context it is worth to mention that vacuum polarization 
for negative $Q^2$ can be obtained directly
using electron-muon scattering process.
In the experiment proposed in ref.~\cite{Abbiendi:2016xup} with 150GeV
muon beam scatering on fixed target electron it will be possible to measure
directly QED effective coupling within the range $0>Q^2 \geq -0.140~{\rm GeV}^2$.
However, SM calculations for FCC-ee processes (low angle Bhabha)
will need vacuum polarization for $|Q^2| \geq 1~{\rm GeV}^2$.

The path of reduction of uncertainties due to pure QED 
corrections was also described in ref.~\cite{Jadach:2018jjo}.
In particular it was pointed out that the dominant \order{\alpha^2 L_e}
correction not included in \bhlumi~4.04 is already known since 
long time ago~\cite{Jadach:1995hy}
and it can be added to the \bhlumi\ matrix element rather easily.
However, the most elegant solution would be to implement in \bhlumi\
the QED matrix element with the same type of the CEEX resummation as in \kkmc,
including in addition the \order{\alpha^3 L_e^3} corrections.
According to ref.~\cite{Jadach:2018jjo}, this would reduce uncertainties due to
all photonic corrections to the level below~$10^{-4}$.

Let us note that subleading ${\cal O}(\alpha^2)$ QED corrections including
non-logarithmic terms to low angle Bhabha
were also calculated in refs.~\cite{Penin:2005eh,Penin:2005kf,Becher:2007cu}.

The quasi-photonic corrections due to emission of the light fermion pairs
are small but difficult to master.
Reduction of their uncertainty below $10^{-4}$ level will require
a special effort and most likely development of another dedicated MC program.
The first step in this direction was done in 
ref.~\cite{Jadach:1996ca,Jadach:1993wk}.

It is quite likely that at the FCC-ee times one may be in the situation
in which the dominant contributions to the luminosity uncertainty $\sim 0.01\%$
will be the experimental one from detector and indirectly the experimental
one from the vacuum polarisation (QCD), 
while the pure QED contribution will be again subdominant.

Summarizing, according to ref.~\cite{Jadach:1995nk},
the total uncertainty of the theoretical prediction for the luminosity cross section
will be at the $10^{-4}$ level, that is a factor $\sim 10$ better that at LEP times.
This will require an upgraded matrix element in \bhlumi\ to \order{\alpha^2} CEEX level,
development of another MC program of comparable class to control technical precision
and further reduction (a factor of 2) of the vacuum polarisation error
contribution.

\subsection{Measurement of $\alpha_{QED}(M_Z)$ using $A_{FB}$ near $Z$ resonance}

The idea of using precision measurement of the charge asymmetry 
$A_{FB}$ near $Z$ resonance in order to extract the QED coupling constant
at the the electroweak scale $\alpha_{QED}(M_Z)$
was proposed quite recently in ref.~\cite{Janot:2015gjr}.
It is an alternative to the current method calculating $\alpha_{QED}(M_Z)$
using as an input very precisely known $\alpha_{QED}(0)$
and the low energy hadronic cross section as an input to dispersion relations.
The new method requires charge asymmetry to be measured with the precision
$\delta A_{FB} \simeq 10^{-5}$ at $\sqrt{s_\pm} \simeq M_Z \pm 3.5$~GeV.
According to ref.~\cite{Janot:2015gjr}
both statistical and experimental systematic error can be
reduced to this fantastic low level, factor $\sim 200$ better than
what we have seen in LEP experimental results.

Concerning QED uncertainties, it was argued in ref.~\cite{Janot:2015gjr}
that because both ISR and IFI corrections do not change sign
across the $Z$ peak, they will cancel in the difference $A_{FB}(s_+)-A_{FB}(s_-)$.
However, in the analysis of ref.~\cite{Jadach:2018lwm}
based on the numerical results of \kkmc\
it was shown that although such a cancellation is present,
nevertheless it is not perfect and
the remaining net effect in the difference is of order $\sim 1\%$.
On the other hand, quite luckily, the QED effects near the the $Z$ resonance,
for a relatively sharp cut-off on the total photon energy  $E_\gamma\leq 0.02~E_{beam}$,
are governed by soft photon emissions.
Because of that, the resummation of higher order corrections
using the soft photon approximation, which is the basis of the CEEX
matrix element of \kkmc, is very efficient,
not only for the total cross section, but also for the charge asymmetry.

In the study of ref.~\cite{Jadach:2018lwm} the results of \kkmc\
for $A_{FB}(s_\pm)$ and their difference were
validated by means of comparing them with the newly
developed MC program \kkfoam\ with partial analytical integration
and soft photon resummation.
It was shown that predictions of \kkmc\ can be trusted
down to the $\delta A_{FB}(s_\pm) \sim 10^{-4}$ precision.
Also, some additional internal tests of \kkmc\ indicate
that the level of  $\delta A_{FB}(s_\pm) \sim 10^{-5}$
is attainable in the results of \kkmc\
for the difference $A_{FB}(s_+)-A_{FB}(s_-)$.
The above results have to be consolidated,
but they indicate that thanks to enormous power
of the soft photon resummation at the amplitude level
developed in ref.~\cite{Jadach:2000ir}
the improvement of the QED prediction for $A_{FB}$
near the $Z$ resonance by factor 200 needed for the FCC-ee experiments
looks feasible.
The above analysis concentrates on the pure photonic
corrections to $A_{FB}$ and many other corrections like
light pair emissions, QCD, electroweak corrections,
in this extreme precision regime will have to be analyzed
very carefully.
In particular it should be checked to what extent
all these corrections/effects cancel in the difference 
$A_{FB}(s_+)-A_{FB}(s_-)$.

Summarizing, the present status of the QED uncertainties (mainly due to IFI)
is that they can be controlled in $A_{FB}(s_+)-A_{FB}(s_-)$ and in 
$\alpha_{QED}(M_Z)$ down to the $\sim 10^{-4}$ level.
Another factor 10 improvement will require hard work but
according to preliminary study in ref.~\cite{Jadach:2000ir} it looks feasible.

\subsection{Invisible $Z$ decay width from cross section and radiative return}
\label{sec:Zinv2}

In the determination of $N_\nu$ from the peak cross section
the error of the luminosity is the main obstacle in the precision improvement.
Using the approximate rule $\delta N_\nu = 7.5\; \frac{\delta {\cal L}}{{\cal L}}$
of ref.~\cite{ALEPH:2005ab}
we estimate that in order to get $\delta N_\nu \simeq 10^{-3}$
the luminosity relative error has to be improved from the present
$\delta {\cal L}/{\cal L} \simeq 5\cdot 10^{-4}$ down to  $1\cdot 10^{-4}$.
As discussed in ref.~\cite{Jadach:2018jjo} it is feasible, 
provided that the QED matrix element in \bhlumi\
is upgraded to the \order{\alpha^2} CEEX level, similarly as in \kkmc,
and the vacuum polarisation at $t=-1GeV^2$ is improved by another factor 2.
The error of  $\sigma(M_Z)$ due to missing higher orders in the QED ISR 
calculations (including light fermion pairs emission)
was conservatively estimated in ref.~\cite{Jadach:1999pp} to be 
$\delta \sigma/\sigma(M_Z) \simeq 2\cdot 10^{-4}$.
Factor 2 improvement is probably obtainable by means of the careful
reexamination of the existing calculations.
Reducing experimental systematic error to the same level is of course
an independent important issue.

The prospects of reducing the QED uncertainty in the
$Z$ radiative return (ZRR) process are discussed in
ref.~\cite{ifjpan-iv-2019xx}.
Here we only summarize the main points.

First of all, 
in the direct determination of $N_\nu$ from the ZRR neutrino-like cross section
one has to take into account that
the QED matrix element of  \kkmc\ is much better than that of
{\tt KORALZ}~\cite{Jadach:1991ws} and {\tt NUNUGPV98}~\cite{Montagna:1998ce} 
used at LEP.
Nevertheless, due to selecting event with
one real photon, one is effectively loosing one perturbative order
and it is effectively ``downgraded'' from to the complete 
${\cal O} (\alpha^2)_{exp.}$ to ${\cal O} (\alpha^1)_{exp.}$.
Validating precision of the  \kkmc\ for the ZRR process is not yet completed --
an optimistic estimate of ref.~\cite{ifjpan-iv-2019xx} for 
the photonic uncertainty of the ZRR process is quite promising,
$\delta N_\nu/N_\nu \simeq 2.4\cdot 10^{-5}$ for 105~GeV%
\footnote{The energy $\sqrt{s}=105$GeV is not included in the FCC-ee operational mode.}
and $\delta N_\nu/N_\nu \simeq 2.2\cdot 10^{-4}$ for 161~GeV.
The present error due to luminosity 
$\delta {\cal L}/{\cal L} \simeq 5\cdot 10^{-4}$ \cite{ALEPH:2005ab} implies
$\delta N_\nu/N_\nu =  3 \frac{\delta {\cal L} }{ {\cal L} } \simeq 1.5\cdot 10^{-3} $
and it would dominate, 
although its reduction by factor 2-5 according
to ref.~\cite{Jadach:2018jjo} is feasible.
More conservatively,
for full exploitation of the FCC-ee precision it would be recommended
to develop a new dedicated MC program with 
the complete ${\cal O}(\alpha^2)_{exp.}$ matrix element for the ZRR neutrino-like process.
In the mean time one should validate more precisely
predictions of \kkmc\ for ZRR by means of comparison with the
semi-analytical programs based on collinear structure functions
similar to \kksem~\cite{Jadach:2000ir} or {\tt NUNUGPV98}\cite{Montagna:1998ce}.

In the method of extracting $N_\nu$ from the ratio of the
ZRR photon spectrum for the invisible $Z$ decay and $Z$ decaying into muon pairs,
the luminosity will cancel out. In addition the effect of the ISR
will also cancel quite precisely as well, as shown in 
ref.~\cite{ifjpan-iv-2019xx}.
On the other hand, due to the cutoff
on the photon angle a sizable QED FSR effect in the muon pair process
will remain in the ratio, see~\cite{ifjpan-iv-2019xx}.
NB. The contribution from the IFI is also sizable, but does not contribute much
to integrated cross section because it changes sign
across the $Z$ peak in the photon spectrum.
Consequently, a high quality MC program for predicting ZRR with muon pairs
will be mandatory.
\kkmc\ will be obviously helpful, provided its validation using
auxiliary programs is pursued, but ultimately a genuine  
${\cal O}(\alpha^2)_{exp.}$ for the muonic ZRR is highly desirable. 

Summarizing, the optimistic estimate of the photonic QED correction
in ref.~\cite{ifjpan-iv-2019xx},
based on the internal crosschecks of \kkmc,
(dominated by the FSR effect in the muon-pair ZRR)
is $\delta N_\nu \simeq 0.9 \cdot 10^{-3}$,
for both c.m.s. energies, 105~GeV and 161~GeV.
It would be necessary to develop a new MC program dedicated to the ZRR
process in order to crosscheck the above result
and gain another factor 3 needed for pushing the uncertainty of QED corrections
to a subdominant level.

\subsection{Cross section near the $WW$ threshold ($M_W$ measurement)}
\label{ssect:WWatFCC}

Until recently there was no quantitative reliable estimate 
of the experimental precision of the $M_W$ measurement at FCC-ee
from the $W$ mass reconstruction using final state quarks and leptons%
\footnote{
  Theory uncertainties in this method were estimated in
  ref.~\cite{Schael:2013ita} to be 5-10~MeV,
  while in ref.~\cite{Jadach:2001cz} it was estimated as $\leq 2$~MeV.
}.
According to recent studies presented 
at the FCC Week 2019~\cite{beguin-fcc-week:2019}
both type of measurements, from final state mass reconstruction
and  from the threshold cross section 
will provide precision within a similar range of 0.28-0.45~MeV.
In the following we shall refer mainly to the simpler measurement
of $M_W$ at FCC-ee from the $W$-pair threshold cross section.

The mass of the  $W$ boson can be determined very precisely
from the value of the total cross section
near the threshold of the $e^+e^-\to W^+W^-$ process,
at the c.m.s. energy where the cross section is most sensitive
to $M_W$, which is $\sim 162.5$~GeV~\cite{Benedikt:2018qee}.
(Similarly, the best c.m.s. energy for $\Gamma_W$ determination is 157.5~GeV.)
Knowing the accuracy of the cross section, the precision of the $W$ mass
determined form the threshold scan
can be determined approximately from the following
relation~\cite{Altarelli:1996gh}:
\begin{align}
\Delta M_W = \Delta\sigma_{WW} \Bigl|\frac{dM_W}{d\sigma_{WW}}\Bigr| 
           = \sqrt{\sigma_{WW}} \Bigl|\frac{dM_W}{d\sigma_{WW}}\Bigr| 
                    \frac{\sqrt{\sigma_{WW}}}{\sqrt{N}},
\end{align}
where $\Delta\sigma_{WW} =  \frac{\sigma_{WW}}{\sqrt{N}}$ is the statistical error.
The  sensitivity factor 
$\sqrt{\sigma_{WW}}\Bigl|\frac{dM_W}{d\sigma_{WW}}\Bigr|$
is estimated in \cite{Altarelli:1996gh} to be 
$0.91 \hbox{GeV}/\sqrt{\hbox{pb}}$ at the $WW$ threshold.

The physics goals of FCC-ee are set to
10/ab and $3\times 10^7$ events at the $WW$  threshold \cite{Tenchini-unpublished}. 
For example, for the cross section $\sigma_{WW}\simeq 3$ pb 
and the accuracy of the total cross section
$ \frac{\Delta\sigma_{WW}}{\sigma_{WW}}=1/\sqrt{3\times 10^7}=0.02\% $
the resulting accuracy of the $W$ mass would be
\begin{align}
\Delta M_{W} = 0.91  \frac{\hbox{GeV}}{\sqrt{ \hbox{pb}}} 
                     \frac{\sqrt{3  \hbox{pb}}}{\sqrt{3\times 10^7}} 
                     \simeq 0.3 \hbox{MeV} = 3.6\times 10^{-6} M_W.
\end{align}
The ultimate goal of the measurement of 
the total cross section at the $WW$ threshold at the FCC-ee 
is the precision of 0.02\%.
This requires at least a factor 10 improvement of the theory precision
for the SM calculation of the $W$-pair production near the threshold.

The first step in this direction is obvious 
-- a MC program with the complete ${\cal O}(\alpha)$ 
EW corrections to $e^+e^-\to 4f$ process must be available. 
This task is definitely doable since the analytical/numerical 
calculations already exist. 
That would consolidate the EW precision at the level of 0.2\%. 
In order to reduce the precision further a ${\cal O}(\alpha^2)$ calculation
for the doubly resonant $e^+e^-\to W^+W^-$ process is needed,
with some clever strategies inspired by EFT near the threshold.

A complete EW ${\cal O}(\alpha^2)$ calculation for the $e^+e^-\to 4f$
and ${\cal O}(\alpha^3)$ for $e^+e^-\to W^+W^-$ process
seem to be very hard to do, 
but one may try to estimate uncertainties in case it is not available.
Firstly, let us try to estimate an error due to
the neglected ${\cal O}(\alpha^3)$ EW corrections. 
There are no tools to determine their actual size. 
However, one can do naive scaling  based on the lower order corrections. 
Namely,  if at 161 GeV ${\cal O}(\alpha)\sim 2\% ~\hbox{of~Born}$ (ISR
excluded) and ${\cal O}(\alpha^2)_{} \leq 0.2\% ~\hbox{of~Born}$,
i.e.\ there is a factor of 10 suppression between them,
then we can assume the same factor of 10 between 
${\cal O}(\alpha^2)$ and ${\cal O}(\alpha^3)$ and we end up at 
${\cal O}(\alpha^3) \leq 0.02\% ~\hbox{of~Born}$, which shows that 
${\cal O}(\alpha^3)$ contribution should be just negligible.

Since \order{\alpha^2} complete calculation for the $e^+e^-\to 4f$ process
is a difficult challenge, one should consider another, simpler scheme. 
This scheme would mimic the approach from LEP2 based on calculating 
separately the \order{\alpha^2} corrections to the production and decay of
$W$-pair, i.e.\ to   $e^+e^-\to W^+W^-$ and then $W\to 2f$ process%
\footnote{
  One could profit from two-loop corrections to the muon decay 
  which are known since long \cite{Czakon:2003wg}.}.
Similarly as at LEP2, but at one order higher,
the above calculations would be combined
with the existing \order{\alpha^1} complete calculation for the $e^+e^-\to 4f$
process.
What can we say about precision of such an approach? 
As earlier, we can only do a naive scaling from the lower order. 
Namely, analyzing the Single-pole (SP) and Double-pole (DP) 
contributions at ${\cal O}(\alpha)$ we find from
\cite{Denner:2005es, Denner:2005fg} that at 161~GeV we can estimate the
${\cal O}(\alpha)_{SP}$ as
\begin{align} &
{\cal O}(\alpha)_{SP} \sim \Bigl(
{\cal O}(\alpha)_{4f} - {\cal O}(\alpha)_{DP}
\Bigr)
\sim 2\% ~\hbox{of Born} ~\sim 8\% ~\hbox{of~}{\cal O}(\alpha)_{DP}.
\end{align}
If we now assume, as argued earlier, that ${\cal O}(\alpha^2)\leq 0.2\%$ of 
Born and we extrapolate from the first order that
${\cal O}(\alpha^2)_{SP} \sim 8\% ~{\cal O}(\alpha^2)$, then we arrive at
${\cal O}(\alpha^2)_{SP} \sim 0.016\%$, within the targeted precision of 0.02\%.

Near the threshold, where the split into DP+SP+NP, using decomposition
into powers of $\Gamma_W/M_W$ becomes inefficient,
one should try to exploit expansion in powers of 
$\beta,~\beta^2=(s-4M_W^2)/4M_W^2$,
as in EFT calculations of Refs.~\cite{Beneke:2007zg,Actis:2008rb},
see also the review in Ref.~\cite{Blondel:2019vdq}, 
but for the traditional diagrammatic calculation within the standard phase space,
which can be implemented within the MC event generator.
This may reduce significantly the number of diagrams in the  
${\cal O}(\alpha^2)$ for $e^+e^-\to 4f$ to a manageable subset.

In subsection~\ref{ssect:WWatLEP} it was underlined that
the implementation of the QED universal corrections 
in \racoon\ on one hand and in \koralw\ and \yfsww3 on the other hand
was very different.
The immediate question is whether these approaches can be easily extended
to new Monte Carlo event generators including
complete \order{\alpha^2} for $e^+e^-\to W^+W^-$ matrix element
or any complete or partly complete \order{\alpha^2} for $e^+e^-\to 4f$?
It looks that it will be very hard to extend the methodology of \racoon\ to 
\order{\alpha^2} --
the subtraction technique of Catani-Seymour is limited to \order{\alpha^1}
and the ''sectorisation'' technique becomes very complicated beyond
\order{\alpha^1}.
On the other hand, an example of the MC, \kkmc\ program,
with complete \order{\alpha^2} QED for the $e^+e^-\to 2f$ process exists already,
and YFS-inspired CEEX technique~\cite{Jadach:1999vf,Jadach:2000ir} 
of factorizing and resumming universal QED correction
within the MC event generator will work at the practical level 
both for $e^+e^-\to W^+W^-$ and $e^+e^-\to 4f$ processes.
In particular, the CEEX technique is very well suited for implementing
''non-factorisable'' interferences between the $W^+W^-$ production process
and two $W$ decays and between the two $W$ decays as well.
It extends the exponentiation of soft photons of ref.~\cite{Yennie:1961ad}. 
In the classical approach  one
exponentiates only interferences and emissions from external legs. 

In the new CEEX-style scheme sketched 
in refs.~\cite{Jadach-Zeuthen:2001,Jadach:2019wol}  
also photon radiation including
interferences from the internal charged W bosons is exponentiated.
One may ask whether it makes sense as there are 
no related infrared singularities present.
The answer is yes, because $W$s can be treated as almost stable particles. 
The main features of the above scheme are the following:
\begin{itemize}
\item
The missing interferences for production-decay and decay-decay to the $WW$ graphs
are included {\em to all orders in soft approximation}.
\item
Shift in kinematics due to interferences/emissions (recoil) in W propagators is 
properly described.
\item
Higher order corrections can be easily added to the non-soft functions%
\footnote{%
 Non-soft functions at \order{\alpha^1} 
 for double-resonant and single-resonat parts of the $e^+e^-\to 4f$ process
 are defined in subsections 4.2 and 4.3 of ref.~\cite{Jadach:2019wol}.
 For non-resonant part they are defined the same way as in the standard YFS/CEEX scheme,
 see also Sect.~C.2.7 in Ref.~\cite{Blondel:2018mad}.
}
resulting from the IR subtractions.
\end{itemize}
The above scheme would work for single-$W$ or $ZZ$ pair production as well.
The implementation may start with simpler EEX-type scheme,
already partly implemented  for $e^+e^-\to W^+W^-$ in \yfsww3
and for $W^\pm$ decays in \winhac\ MC~\cite{Placzek:2003zg},
without production-decay and decay-decay interferences.
The CEEX scheme would be implemented in the next
step by means of reweighting MC events with
multiplicative weight, similarly as it is done in \kkmc.
The above would take care of the doubly resonant part of $e^+e^-\to 4f$.
Implementation of the CEEX matrix element for
the single-resonant part would be straightforward.
For the non-resonant part one probably needs only CEEX for ISR
and some crude FSR implementation, neglecting IFF interferences.

Summarizing,
the precision of the $WW$  legacy MC codes from LEP2 is 2\% at the threshold, 
as follows from the direct comparison with the complete calculation 
\cite{Denner:2005es, Denner:2005fg}. 
The inclusion of this complete ${\cal O}(\alpha)$ calculation would 
improve EW precision to the level of 0.2\%, 
as follows from the dominant ${\cal O}(\alpha^2)$ calculation of 
\cite{Beneke:2007zg,Actis:2008rb} 
and estimates in \cite{Denner:2005es, Denner:2005fg}. 
To achieve 0.02\% EW precision of the cross-section one has to compute 
and implement the ${\cal O}(\alpha^2)$ EW corrections. 
We argued that it could be enough to calculate them in 
double/leading pole approximation,
supported by the methods of near-threshold-improvements of
refs.~\cite{Beneke:2007zg,Actis:2008rb}.
The exponentiation of real and virtual radiation from intermediate $W$s
proposed in ref.~\cite{Jadach-Zeuthen:2001} can further improve the precision.
The QCD corrections have to be analyzed and included as well.
This way the overall SM precision tag $\sim 10^{-4}$ 
for the $e^+e^-\to W^+W^-$ process seems feasible.
The \yfsww+\koralw\ approach looks like a good starting point for
the above developments.
The same improvements of the theoretical calculations and MC programs
are essential for $M_W$ measurements using final state mass reconstruction.

\subsection{New ideas on pseudo-observables at FCC-ee}
\label{ssect:ewpos}

\begin{figure}[t]
  \centering
  \includegraphics[width=130mm]{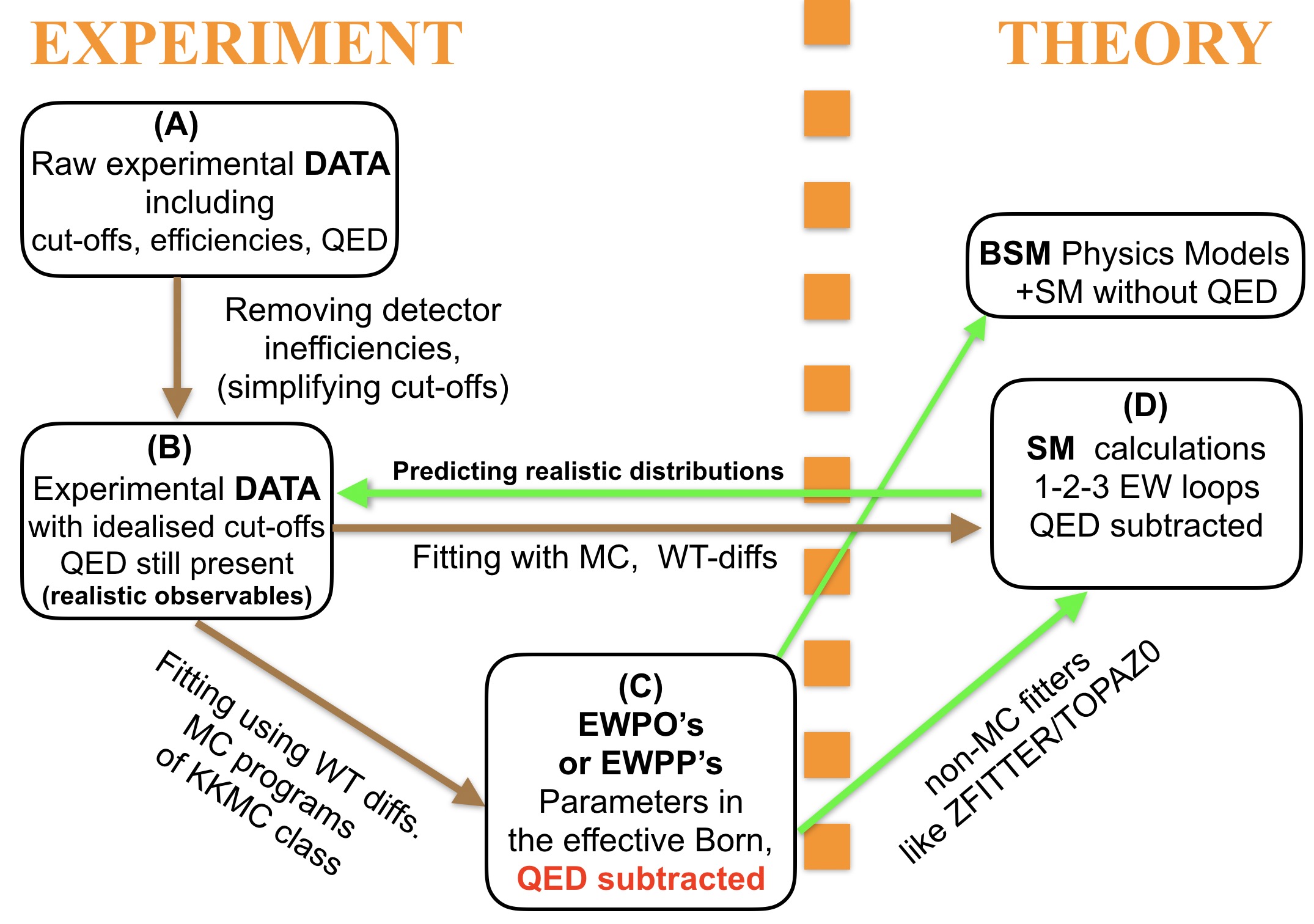}
  \caption{\sf
  Scheme of construction and the use of EWPO/EWPP at FCC-ee.
  }
  \label{fig:cEWPO}
\end{figure}

The system of electroweak pseudo-observable (EWPOs) used in final
analysis of LEP data near $Z$ resonance
was forged in the interaction between theorists
and experimentalists~\cite{Bardin:1999gt,ALEPH:2005ab}.
As already said, it may not work at the higher experimental precision of FCC-ee.
The authors of Section C2 of ref.\cite{Blondel:2018mad}
are proposing to modify the EWPO scheme of LEP to the FCC-ee level.
The main aim is to preserve its fundamental feature,
that is to provide a flexible link between data and theory.
EWPOs should provide model independent encapsulation 
(representation) of experimental data,
where model independence means removing from data technical details
of the detector (inefficiencies), kinematic cut-offs 
and large universal QED corrections.

Let us explain this new modified scheme of EWPOs
following its graphical representation in Fig.~\ref{fig:cEWPO}.
Similarly as in the LEP case, the EWPOs are encoded in the parameters of some
effective Born (couplings, masses) such that EWPOs like cross sections,
asymmetries, partial widths are in one-to-one correspondence with
these parameters, which are called EW pseudo-parameters (EWPPs).
The critical question is whether one may factorize-off and remove
QED corrections at the FCC-ee precision level on the way
from raw experimental data to EWPOs/EWPPs, $(A)\to(C)$ in Fig.~\ref{fig:cEWPO}?

As indicated in the previous section, the QED deconvolution of LEP,
which was done at the amplitude squared level and ignoring non-factorisable IFI effects,
will be unable to meet FCC-ee precision criteria.
However, the QED factorisation at the amplitude level has the extra power for the task.
The price to pay is that it cannot be done analytically,
but has to be implemented using numerical sesummation within the Monte Carlo.
The corresponding transformation of data $(B)\to(C)$ in Fig.~\ref{fig:cEWPO}
was done in the LEP scenario using \zfitter/\topazz\ programs and in
the proposed scheme the fitting
parameters in the effective Born would be done by the Monte Carlo programs.
An additional advantage is the full flexibility concerning 
the choice of experimental cut-offs at the stages $(A)$ or $(B)$ in 
Fig.~\ref{fig:cEWPO}.
In addition, the MC program will have not only effective Born amplitudes
convoluted with multiple photon emissions, but also \order{\alpha^1}
or even \order{\alpha^2} non-IR electroweak corrections,
thus it will be able to fit directly data to SM predictions,
as indicated in $(B)\to(D)$ in Fig.~\ref{fig:cEWPO}.
This is important, because what we want for new EWPOs/EWPPs is possibility
of fitting data to SM, $(C)\to(D)$ in Fig.~\ref{fig:cEWPO},
as in case of LEP scenario.

Direct fit of data to the SM predictions, $(B)\to(D)$, 
will provide a crosscheck on the precision loss 
in the desired two-step scenario $(B)\to(C)\to(D)$.
Another powerful crosscheck done in ref.~\cite{Bardin:1999gt}
was to close the loop $(B)\to(C)\to(D)\to(B)$.
This kind of cross-checks were also done in ref.~\cite{ALEPH:2005ab} 
in the final LEP data analysis.
Obviously, the role of the MC in the proposed scheme 
is more important than at LEP scenario.
It will also require implementing in the MC programs certain
provisions for fast evaluation of the changes of the 
SM predictions due to small variation of the SM input parameters,
using the technique of weight differences.

It should be stressed that scenario proposed above is one of many possibilitities.
It will be the aim of the future studies and practice of the FCC-ee data analysis
to decide about its final shape.
For instance, in the above LEP-like scenario the main objective 
in the $(B)\to(C)$ transition is to encapsulate
experimental data in the form of Born like parametrization of the vector 
couplings of the $Z$ exchange,
while background non-resonant exchange is parametrized 
in a rather simple way as photon exchange.
One may consider departing from the above by means of 
including in the simplified spin amplitudes used in $(B)\to(C)$ stage
most of the \order{\alpha^1} EW corrections (like EW boxes),
such that they are removed from data.
In this way EWPOs/EWPPs would represent more cleanly possible signals of new physics.
One may also think about extending list of EWPPs with parameters
residing in the non-resonant part of spin amplitudes.
This would help to describe them in a refined way,
again with the aim of exposing more efficiently
possible signals of new physics in experimental data.

The above analysis is only indicating certain possible avenues
of reformulating the methodology of electroweak pseudo-observables 
and more detailed analysis is badly needed.
In the end, experimentalists in close collaboration with theorists
will decide on how FCC-ee data will be  analyzed.

\begin{table}
\centering
\begin{tabular}{|c|c|c|c|c|l|}
\hline
 Observable & Source & Err.\{QED\}
                     & Stat[Syst]& LEP  & main development \\
            & LEP    & LEP
                     & FCC-ee    &  $\overline{\text{FCC-ee}}$ &  to be done\\
\hline
$M_Z$ [MeV]    & $Z$ linesh.
               & $2.1\{0.3\}$                  & $0.005 [ 0.1 ]$ 
               & 3$\times 3^{\star}$ & light fermion pairs\\
$\Gamma_Z$ [MeV] & $Z$ linesh.
               &  $2.1\{0.2\}$                 & $0.008[ 0.1 ] $ 
               & 2$\times 3^{\star}$ &  fermion pairs\\
$R^Z_l \times 10^{3} $ & $\sigma(M_Z)$
               & $25\{12\}$                    & $ 0.06[ 1.0 ] $  
               & 12$\times 3^{\star\star}$ & better FSR\\
$\sigma^0_{\rm had}$ [pb] & $\sigma^0_{\rm had}$
               & $37\{25 \}$                   & $ 0.1[ 4.0 ]$    
               & 6$\times 3^{\star}$ & better lumi MC\\
$N_\nu \times 10^{3}$      & $\sigma(M_Z)$
               & $8\{6\} $                     &$ 0.005[ 1.0 ]  $ 
               & 6$\times 3^{\star\star}$ & CEEX in lumi MC \\
$N_\nu \times 10^{3}$      & $Z\gamma$
               & $150\{60\} $                  & $ 0.8[ <1 ]  $
               & 60$\times 3^{\star\star}$ & \order{\alpha^2} for $Z\gamma$\\
$\sin^2\theta_W^{eff}\times 10^{5}$ & $A_{FB}^{lept.}$
               & $53\{28\}$                   & $ 0.3[ 0.5 ] $ 
               & 55$\times 3^{\star\star}$  & h.o. and EWPOs \\
$\sin^2\theta_W^{eff}\times 10^{5}$      &$\langle{\cal P}_\tau\rangle$,$A_{\rm 
FB}^{pol,\tau}$
               & $41\{12\}$                  & $0.6[ <0.6 ] $ 
               & 20$\times 3^{\star\star}$  & better $\tau$ decay MC\\
$M_W$ [MeV]    & mass rec.
               & $33\{6\}$                   &   $0.5[ 0.3 ]$  
               & 12$\times 3^{\star\star\star}$  & QED at threshold\\
 {\small $A_{FB,\mu}^{M_Z\pm 3.5 {\rm GeV}}\!\!\!\times 10^{5} $}
 & $\frac{d\sigma}{d\cos\theta}$
               & $2000\{100\}$               & $1.0[ 0.3 ]$      
               & 100$\times 3^{\star\star\star}$  & improved IFI\\

\hline
\end{tabular}
\caption{\sf
 Comparing experimental and theoretical errors at LEP and FCC-ee as in 
 Table~\ref{tab:table1}.
 3rd column shows LEP experimental error together with uncertainty induced by QED
 and 4th column shows anticipated FCC-ee experimental statistical [systematic] errors.
 Additional factor $\times 3$ in the 5-th column (4th in Table~\ref{tab:table1})
 reflects what is needed for QED effects to be {\em subdominant}.
 Rating from $^\star$  to $^{\star\star\star}$ marks whether the needed improvement
 is relatively straightforward, 
 difficult or very difficult to achieve.
}
\label{tab:table2}
\end{table}

\section{Summary}

The main results of our study are indicated in Table~\ref{tab:table2},
where we have indicated for selected observables, 
the same as in Table~\ref{tab:table1},
the improvement factor needed in calculations of the QED effects in order
to match the experimental precision anticipated in the FCC-ee experiment.
We have also indicated explicitly the additional factor 3 necessary for
these effects to become subdominant
and the most important development to be done.
We have also tried to rate how difficult it will be to achieve these targets.
It is needless to say that many of the above estimates remain speculative
and are not based on solid numerical results.
However, this is always the case with estimates of the uncalculated 
higher order perturbative corrections,
so they have to be always taken with a grain of salt.
On the other hand, this kind of analysis is indispensable
in planning directions and priorities of the future work.

\vspace{3mm}
{\bf\large Acknowledgments}

Authors would like to thank warmly 
J.~Gluza, A.~Freitas, P.~Janot, W.~P\l{}aczek, Z.~W\c{a}s and B.F.L~Ward for reading 
the manuscript critically.



\end{document}